\documentstyle[aps,pra]{revtex}
\draft
\begin{document}
\tightenlines

\title{Landau-Khalatnikov two-fluid hydrodynamics
 of a trapped Bose gas}

\author{Tetsuro Nikuni and Allan Griffin}
\address{Department of Physics, University of Toronto,
Toronto, Ontario, Canada M5S 1A7} 

\date{\today}
\maketitle

\begin{abstract}
Starting from the quantum kinetic equation for the non-condensate atoms
and the generalized Gross-Pitaevskii equation for the condensate,
we derive the two-fluid hydrodynamic equations of a trapped Bose gas
at finite temperatures.
We follow the standard
Chapman-Enskog procedure, starting from a solution of the kinetic
equation corresponding to the complete local
equilibrium between the condensate and the non-condensate components.
Our hydrodynamic equations are shown to reduce to a form identical 
to the well-known Landau-Khalatnikov two-fluid equations,
with hydrodynamic damping due to the deviation from local equilibrium.
The deviation from local equilibrium  within the thermal cloud
gives rise to dissipation associated with shear viscosity and thermal conduction.
In addition, we show that effects due to the
deviation from the diffusive local equilibrium between the
condensate and the non-condensate (recently considered by Zaremba, Nikuni
and Griffin) can be described by four
frequency-dependent second viscosity transport coefficients.
We also derive explicit formulas for all the transport coefficients.
These results are used to introduce two new characteristic relaxation
times associated with hydrodynamic damping.
These relaxation times give the rate at which local equilibrium is reached
and hence determine whether one is in the two-fluid hydrodynamic region.

\end{abstract}

\section{introduction}
\label{sec:intro}
At very low temperatures, the dynamics of a trapped Bose gas is described by
the time-dependent Gross-Pitaevskii (GP) equation for the macroscopic
wavefunction of the condensate. 
As discussed in several
recent reviews \cite{RMP,varenna}, there is excellent agreement
(within a few percent) between experimental observations of 
collective modes for $T\lesssim 0.4T_{\rm BEC}$ 
and theoretical calculations based on the $T=0$ GP equation.
At elevated temperatures where the condensate is appreciably depleted by
thermal excitations, one must consider the coupled motion of the
condensate and non-condensate degrees of freedom.
In a recent paper, Zaremba, Nikuni and Griffin \cite{ZNG} derived a generalized Gross-Pitaevskii
equation for the condensate atoms and a quantum kinetic equation for
the non-condensate atoms, which can be used to discuss the coupled
dynamics of the two components at finite temperatures. 
These two components are coupled through mean-field interactions as well
as collisions between the atoms.

Two limiting cases for
the dynamics of the gas correspond to the collisionless and hydrodynamic
regimes~\cite{varenna,NP}.  
Up to the present, most experiments on the collective modes of Bose
gases are thought to be in the low-density collisionless limit.
In this regime, the main 
effect of the non-condensate (thermal cloud) component 
is to damp the condensate oscillations.
In addition to Landau and Beliaev damping
due to the dynamic mean-field interaction between two components,
there is damping arising from the lack of diffusive local equilibrium between
the condensate and the non-condensate \cite{griffin}.
The latter mechanism of damping has recently been worked out in detail by
Williams and Griffin \cite{WG}.

In contrast, in the regime where
collisions between atoms are rapid enough to establish a state of 
dynamic local equilibrium in the non-condensate gas,
the dynamics of the system is described by hydrodynamic equations for
a few local variables.
This regime contains much new physics and it should be accessible taking
advantage of the larger densities now available as well as the large 
scattering cross sections close to a Feshbach resonance~\cite{85Rb}.
We have recently given a detailed derivation and discussion of two-fluid
hydrodynamic equations for trapped atomic gases at finite
temperatures~\cite{ZNG,griffin,ZGN,NZG,CJP},
starting from a generalized GP equation coupled with a kinetic equation.
These equations were derived under the assumption that
the non-condensate atoms are in local thermodynamic equilibrium 
among themselves but are {\it not}
in diffusive equilibrium with the condensate atoms.
The resulting ZGN$'$ hydrodynamic equations~\cite{NZG,ZNG} involve a
new characteristic relaxation time $\tau_{\mu}$, which is the time scale on
which local diffusive equilibrium is established. 
This equilibration
process leads to a novel damping mechanism which is associated with
the collisional exchange of atoms between the two components.
This ZGN$'$ hydrodynamics is briefly reviewed in Section \ref{zgn'}.
In Ref.~\cite{CJP}, we generalized the ZGN$'$ equations to include the effect of
deviations from local equilibrium, and worked out hydrodynamic damping
associated with the collisions among the non-condensate atoms.
At finite temperatures of interest, this deviation from local equilibrium
within the thermal cloud gives rise to damping associated with
thermal conductivity and the shear viscosity.
Such a generalization was first discussed in Section V of Ref.~\cite{NG} starting from
the ZGN hydrodynamic equations.
We also note that the thermal conductivity and shear viscosity were first derived for a
{\it uniform} Bose-condensed gas at finite temperatures in pioneering papers by Kirkpatrick
and Dorfman~\cite{KD}.

In the present paper, building on our recent work with Zaremba, we present
a more complete derivation of two-fluid hydrodynamic equations,
including dissipation.
In Section \ref{sec:CE},
we solve the kinetic equation by expanding 
the distribution function $f({\bf r},{\bf p},t)$ around $f^{(0)}({\bf r},{\bf p},t)$,
the latter describing
complete local equilibrium between the condensate and the non-condensate.
We follow the standard Chapman-Enskog method used to derive hydrodynamic damping
in the kinetic theory of classical gases.
In this treatment, the lowest order hydrodynamic equations involve no
dissipative terms.
All hydrodynamic damping effects are included by taking into account deviations
from the local equilibrium distribution $f^{(0)}$.
In Section \ref{sec:LK}, we prove that,
with appropriate definitions of various thermodynamic variables, 
our two-fluid hydrodynamic equations with damping have precisely the structure of
those first derived by Landau and Khalatnikov~\cite{Landau,Khal}. 
In particular, the damping associated with
the collisional exchange of atoms between the two components, which has been discussed
at length in our previous work \cite{ZNG,NZG,CJP}, is now expressed in terms of
{\it frequency-dependent} second viscosity coefficients.
This type of damping is in addition to the usual kind of hydrodynamic damping associated
with shear viscosity and thermal conductivity \cite{CJP,NG}.

In Section \ref{sec:transport}, we also derive explicit expressions for
all the transport coefficients that
appear in the dissipative terms in our two-fluid hydrodynamic equations. 
For the purpose of illustration, we evaluate the temperature dependence of all
the transport coefficients for the case of a {\it uniform} Bose gas.
The case of a trapped Bose gas is quite different.
Because the condensate density is always much larger than the non-condensate
density in the central regions of a trapped Bose gas, we find that the collisions
between the condensate and non-condensate atoms are the dominant contribution
to all transport coefficients.

The lengthy analysis in Sections \ref{sec:CE}, \ref{sec:LK} and \ref{sec:transport}
is, of necessity, very complex and conceptionally quite subtle.
Most readers will only be interested in the final conclusions, which we now
summarize.
We prove that the two-fluid hydrodynamic equations of a trapped Bose gas can be written
precisely in the well-known Landau-Khalatnikov form, as summarized in
(\ref{landau_eqs}).
The new feature which arises in a trapped Bose gas (as compared to 
superfluid $^4$He) is that, as noted above, the four second viscosity
coefficients can be frequency-dependent, as given by (\ref{zeta_omega}).
Finally, explicit expressions for all the transport coefficients are given in
(\ref{kappa_final}), (\ref{etafinal}) and (\ref{zeta_final}).
An important final result of our analysis (see Section \ref{sec:transport}) is a
precise definition of three characteristic relaxation times associated with the
various transport coefficients.
Moreover we show that the collisions between the condensate and non-condensate
atoms always play the dominant role in the hydrodynamic damping of trapped Bose gases.

The present paper brings to a natural conclusion our series of papers (with Zaremba)
on the two-fluid hydrodynamics of trapped Bose gases \cite{ZNG,NZG,CJP,NG}.
Much remains to be done solving these hydrodynamic equations and experimentally
checking the predictions.

\section{ZGN$'$ hydrodynamic equations}
\label{zgn'}
In this section, we derive the most general form of hydrodynamic equations
for the condensate and non-condensate at finite temperatures,
and then briefly review the ZGN$'$ two-fluid hydrodynamics.
We start with the underlying coupled dynamical equations for the non-condensate
and the condensate, as derived in Ref.~\cite{ZNG} and recently
reviewed in Ref.~\cite{griffin}.
The non-condensate atoms are described by the distribution function
$f({\bf r},{\bf p},t)$, which obeys the quantum kinetic equation
\begin{eqnarray}
{\partial f({\bf r},{\bf p},t) \over \partial t} + {{\bf p} \over m} 
\cdot \bbox{\nabla}_{{\bf r}} f({\bf r},{\bf p},t) &-& \bbox{\nabla}_{{\bf r}}
U \cdot \bbox{\nabla}_{{\bf p}} f({\bf r} ,{\bf p},t) \cr
&=& C_{12}[f,\Phi] + C_{22}[f].
\label{eq1}
\end{eqnarray}
Here the effective potential
$U({\bf r},t)\equiv U_{\rm ext}({\bf r})+2g[n_c({\bf r},t)+\tilde 
n({\bf r},t)]$ includes the self-consistent Hartree-Fock (HF) mean field, and as usual,
we treat the inter-atomic interaction in the $s$-wave approximation with
$g=4\pi \hbar^2 a/m$. 
The condensate density is $n_c({\bf r},t)\equiv|\Phi({\bf r},t)|^2$ 
and the non-condensate density $\tilde n({\bf r},t)$ is given by
\begin{equation}
\tilde n({\bf r},t)=\int\frac{d{\bf p}}{(2\pi\hbar)^3}f({\bf r},{\bf p},t).
\label{eq2}
\end{equation}
The two collision terms in (\ref{eq1}) are given by
\begin{eqnarray}
&&C_{22}[f] \equiv \frac{2 g^2}{(2\pi)^5\hbar^7}
 \int{d{\bf p}_2}\int{d{\bf p}_3}
\int d{\bf p}_4 \cr
&&\times\delta ({\bf p}+{\bf p}_2 -{\bf p}_3 -{\bf p}_4)
\delta(\tilde\varepsilon_{p}+\tilde\varepsilon_{p_2}
-\tilde\varepsilon_{p_3}-\tilde\varepsilon_{p_4}) \cr
&&\times\left[(1+f)(1+f_2)f_3f_4-ff_2(1+f_3)(1+f_4)\right]\, ,
\label{eq3}
\end{eqnarray}
\begin{eqnarray}
&&C_{12}[f,\Phi]\equiv \frac{2 g^2 n_c}{(2\pi)^2\hbar^4} \int d{\bf p}_1
\int d{\bf p}_2 \int d{\bf p}_3 \cr
&&\times  \delta(m{\bf v}_c+{\bf p}_1-{\bf p}_2-{\bf p}_3)
\delta(\varepsilon_c+\tilde\varepsilon_{p_1}
-\tilde\varepsilon_{p_2}-\tilde\varepsilon_{p_3}) \cr
&&\times  [\delta({\bf p}-{\bf p}_1)-\delta({\bf p}-{\bf p}_2)
-\delta({\bf p}-{\bf p}_3)] \cr
&&\times [(1+f_1)f_2f_3-f_1(1+f_2)(1+f_3)],
\label{eq4}
\end{eqnarray}
with $f \equiv f({\bf r, p}, t),\, f_i\equiv f({\bf r, p}_i, t)$.
The expression in (\ref{eq4}) takes into account the fact that a
condensate atom locally has energy
$\varepsilon_c({\bf r},t)=\mu_c({\bf r},t)+\frac{1}{2}mv_c^2({\bf r},t)$
 and momentum $m{\bf v}_c$, where the condensate chemical potential 
$\mu_c$ and velocity ${\bf v}_c$ will be defined shortly.
On the other hand, in our finite-temperature model,
a non-condensate atom locally has the HF energy
$\tilde\varepsilon_p({\bf r},t)=\frac{p^2}{2m}+U({\bf r},t)$.
This particle-like dispersion relation limits our entire analysis 
to finite temperatures.

The equation of motion for the condensate
is given by a generalized 
Gross-Pitaevskii equation \cite{ZNG} for the macroscopic wavefunction 
$\Phi({\bf r},t)$ (see also Ref.~\cite{stoof})
\begin{equation}
i\hbar \frac{\partial\Phi({\bf r},t)}{\partial t}=
\left[-\frac{\hbar^2\nabla^2}{2m}+U_{\rm ext}({\bf r})
+gn_c({\bf r},t)+2g\tilde n({\bf r},t)-i R({\bf r},t)\right]\Phi({\bf r},t),
\end{equation}
where
\begin{equation}
R({\bf r},t)=\frac{\hbar\Gamma_{12}({\bf r},t)}{2n_c({\bf r},t)},
\end{equation}
with
\begin{equation}
\Gamma_{12}\equiv\int\frac{d{\bf p}}{(2\pi \hbar )^3}C_{12}[f({\bf r},{\bf p},t),\Phi({\bf r},t)].
\label{def_G12}
\end{equation}
The dissipative term $R$ in (5) is associated with the exchange of atoms
between the condensate and non-condensate, as described by the collision
integral $C_{12}[f,\Phi]$ in (4).
We see that (1) and (5) must be solved self-consistently.
It is customary to rewrite the GP equation (5) in terms of the amplitude and phase of
$\Phi({\bf r},t)=\sqrt{n_c({\bf r},t)}e^{i\theta({\bf r},t)}$, which leads to
(${\bf v}_c\equiv\hbar \bbox{\nabla}\theta({\bf r},t)/m$)
\begin{mathletters}
\begin{eqnarray}
{\partial n_c \over \partial t} + \bbox{\nabla}\cdot(n_c{\bf v}_c)&=& 
-\Gamma_{12}[f,\Phi]\,,  
\label{eq_nc}
\\
m\left({\partial\over\partial t}+{\bf v}_c\cdot 
\bbox{\nabla}\right) {\bf v}_c&=&-\bbox{\nabla}\mu_c \ ,
\label{eq_vc}
\end{eqnarray}
\label{hydro-C}
\end{mathletters}

\noindent
where the condensate chemical potential is given by
\begin{equation}
\mu_c({\bf r}, t) =-\frac{\hbar^2\nabla^2\sqrt{n_c({\bf r},t)} }{2m\sqrt{n_c({\bf r},t)}}
+U_{\rm ext}({\bf r})+
gn_c({\bf r}, t)+2g\tilde{n}({\bf r}, t)\, .
\label{mu_c}
\end{equation}
One sees that $\Gamma_{12}$ in (8) plays the role of a ``source function" in the 
continuity equation for the condensate, arising from the fact that $C_{12}$ collisions
do not conserve the number of condensate atoms \cite{ZNG}.

We note that the set of equations (\ref{eq1})-(\ref{def_G12}) has also been derived
using the elegant Kadanoff-Baym formulation \cite{milena1,milena2,KB}.
More recently, this KB derivation has been extended to cover low temperatures
as well \cite{milena3}, by working with a Bogoliubov quasiparticle
spectrum instead of the simpler HF spectrum used in the present paper.
One could use this extension as the basis for generalizing the present paper.

Following the standard procedure in the classical kinetic theory of gases \cite{huang}, 
we take moments of the kinetic equation (\ref{eq1}) with respect to $1,{\bf p}$ and $p^2$
to derive the most general form of ``hydrodynamic equations" for the
non-condensate. These moment equations take the form ($\mu$ and $\nu$ 
are Cartesian components):
\begin{mathletters}
\begin{eqnarray}
&&{\partial{\tilde n}\over\partial t}+\bbox{\nabla}\cdot 
(\tilde{n}{\bf v}_n) = \Gamma_{12}[f]\,, \\
&&m{\tilde n}\left({\partial\over\partial t}+{\bf v}_n\cdot 
\bbox{\nabla}\right) v_{n\mu}=-{\partial P_{\mu\nu}\over\partial x_\nu}
-{\tilde n}{\partial U\over\partial x_\mu}
-m(v_{n\mu}-v_{c\mu})\Gamma_{12}[f]\,, 
\\
&&{\partial\tilde\epsilon\over\partial t} +
\nabla\cdot(\tilde\epsilon{\bf v}_n) = -\bbox{\nabla}\cdot{\bf Q}
-D_{\mu\nu} P_{\mu\nu}+ \left[\frac{1}{2}m({\bf v}_n-{\bf v}_c)^2
+\mu_c-U\right]\Gamma_{12}[f]. 
\end{eqnarray}
\label{hydro_general}
\end{mathletters}

\noindent
The non-condensate density was defined earlier in (\ref{eq2}), 
while the non-condensate local velocity is defined by
\begin{equation}
{\tilde n}({\bf r},t){\bf v}_n({\bf r}, t)\equiv\int{d{\bf p}
\over(2\pi\hbar)^3} {{\bf p}\over m} f({\bf r, p}, t)\,.
\label{vn}
\end{equation}
In addition, we have introduced the following quantities:
\begin{mathletters}
\begin{eqnarray}
P_{\mu\nu}({\bf r}, t)&\equiv& m
\int{d{\bf p}\over(2\pi\hbar)^3}\left({p_\mu\over m} - v_{n\mu}\right)
\left({p_\nu\over m} - v_{n\nu}\right)
f({\bf r, p}, t), \label{eq35a} \\
{\bf Q}({\bf r}, t)&\equiv& \int{d{\bf p}\over(2\pi\hbar)^3} {1\over 2m} 
({\bf p}-m{\bf v}_n)^2\left({{\bf p}\over m}-{\bf v}_n\right)
f({\bf r, p}, t),\label{eq35b}  \\
\tilde \epsilon({\bf r}, t) &\equiv&\int{d{\bf p}\over(2\pi\hbar)^3}
{1\over 2m} ({\bf p}-m{\bf v}_n)^2
f({\bf r},  {\bf p}, t) \,. 
\end{eqnarray}
\label{eq35}
\end{mathletters}
Finally, the symmetric rate-of-strain tensor appearing in (10)
is defined as
\begin{equation}
D_{\mu\nu}({\bf r}, t) \equiv {1 \over 2} \left({\partial v_{n \mu}
\over \partial x_\nu} + {\partial v_{n \nu} \over \partial x_\mu}
\right).
\label{eq36}
\end{equation}

The hydrodynamic equations in (\ref{hydro_general}) are formally exact,
but obviously are not closed.
To proceed, we must specify the conditions under which the dynamics of
the system to be determined.
In the ZGN$'$ hydrodynamics, we
consider the situation in which the $C_{22}$ collisions are
sufficiently rapid to establish local equilibrium among the non-condensate atoms.
This situation is described by the local equilibrium Bose distribution
for the thermal cloud, 
\begin{equation}
\tilde f({\bf r},{\bf p},t)=
\frac{1}{e^{\beta[\frac{1}{2m}({\bf p}-m{\bf v}_n)^2+U-\tilde \mu]}-1}
\, .
\label{eq8}
\end{equation}
Here, the temperature parameter $\beta$, normal fluid velocity 
${\bf v}_n$, chemical potential $\tilde \mu$, and mean field $U$ are all
functions of ${\bf r}$ and $t$. 
One may immediately verify that $\tilde f$ has precisely the form such that
it satisfies the condition $C_{22}[\tilde f]=0$,
independent of the value of $\tilde \mu$.  In contrast, 
one finds that $C_{12}[\tilde f]$ does not vanish in general, namely
\begin{eqnarray}
C_{12}[\tilde f]&=&\frac{2g^2 n_c}{(2\pi)^2\hbar^4} 
[1 - e^{-\beta(\tilde \mu-\frac{1}{2}m({\bf v}_n-{\bf v}_c)^2-\mu_c)}] 
\cr && \times \int d{\bf p}_1
\int d{\bf p}_2
\int d{\bf p}_3
\delta(m{\bf v}_c+{\bf p}_1-{\bf p}_2-{\bf p}_3)
\delta(\tilde\varepsilon_1+\varepsilon_c
-\tilde\varepsilon_2-\tilde\varepsilon_3) \cr
&& \times [\delta({\bf p}-{\bf p}_1)-\delta({\bf p}-{\bf p}_2)
-\delta({\bf p}-{\bf p}_3)]
(1+\tilde f_1)\tilde f_2 \tilde f_3.
\label{eq40}
\end{eqnarray}
Using the local distribution function (14) 
to evaluate the moments in (2) and (12), we find that 
the heat current ${\bf Q}({\bf r},t) = 0$, and that
\begin{equation}
\tilde n({\bf r},t) = \int{d{\bf p}\over (2\pi\hbar)^3}
\tilde f({\bf r, p}, t)
= {1\over\Lambda^3} g_{3/2}(z)\,,
\label{eq46}
\end{equation}
\begin{equation}
P_{\mu\nu} ({\bf r}, t) = \delta_{\mu\nu}{\tilde P}({\bf r}, t)
\equiv\delta_{\mu\nu}\int{d{\bf p}\over (2\pi\hbar)^3} 
\frac{({\bf p}^m{\bf v}_n)^2}{3m}
\tilde f({\bf r, p}, t)
=\delta_{\mu\nu}\frac{1}{\beta\Lambda^3}g_{5/2}(z).
\label{eq41}
\end{equation}
Here $z({\bf r},t)\equiv e^{\beta[\tilde\mu({\bf r},t)-U({\bf r},t)]}$ is the local
fugacity, $\Lambda({\bf r},t) \equiv[{2\pi\hbar^2/ mk_B T({\bf r},t)}]^{1/2}$
is the local thermal de Broglie wavelength and 
$g_n(z)=\sum_{l=1}^{\infty}z^l/l^n$ are the Bose-Einstein functions.
The kinetic energy density  is given by
$\tilde \epsilon({\bf r},t) ={3\over 2} \tilde P ({\bf r},t)$.

To summarize, using $f \simeq \tilde f$, we obtain the ZGN$'$ 
lowest-order hydrodynamic equations for the non-condensate given in 
Refs.~\cite{ZNG,griffin,NZG}.
In the linearized version of these ZGN$'$ hydrodynamic equations,
the condensate is described by
\begin{mathletters}
\begin{eqnarray}
&&\frac{\partial \delta n_c}{\partial t}+\bbox{\nabla}\cdot 
(n_{c0}\delta{\bf v_c}) = -\delta \Gamma_{12} \,, \\
&& m\frac{\partial\delta{\bf v}_c}{\partial t}=-\bbox{\nabla}\delta\mu_c,
\end{eqnarray}
\label{lin_C}
\end{mathletters}
and the non-condensate variables satisfy
\begin{mathletters}
\begin{eqnarray}
&&{\partial{\delta \tilde n}\over\partial t}+\bbox{\nabla}\cdot 
(\tilde{n}_0\delta{\bf v}_n) = \delta\Gamma_{12}\,, \\
&&m{\tilde n_0}\frac{\partial \delta{\bf v}_n}{\partial t}
=-\bbox{\nabla} \delta\tilde P
-\delta{\tilde n}\bbox{\nabla} U_0-2g\tilde n_0\bbox{\nabla}(\delta \tilde n
+\delta n_c) \,, \\
&&{\partial\delta\tilde P\over\partial t} =
-\frac{5}{3}\nabla\cdot(\tilde P_0\delta{\bf v}_n) +{2\over 3}
{\bf v}_n \cdot \bbox{\nabla}\tilde P_0 
- {2\over 3}gn_{c0}\delta\Gamma_{12}.
\end{eqnarray}
\label{linZGN'}
\end{mathletters}

\noindent
The fluctuation of the condensate chemical potential is given by 
\begin{equation}
\delta \mu_c=g\delta n_c+2g\delta \tilde n.
\end{equation}
This assumes the Thomas-Fermi approximation, which means that the
first term in (\ref{mu_c}), the quantum pressure term, is neglected.
The source function $\delta\Gamma_{12}$ can be usefully expressed \cite{ZNG}
in terms of the difference between the condensate and non-condensate chemical potentials,
namely
\begin{equation}
\delta \Gamma_{12}=-\frac{\beta_0n_{c0}}{\tau_{12}}(\delta\tilde\mu
-\delta\mu_c)\equiv-\frac{\beta_0n_{c0}}{\tau_{12}}\delta\mu_{\rm diff},
\label{G12}
\end{equation}
where $\mu_{\rm diff}\equiv \tilde\mu-\mu_c$ and 
$\tau_{12}$ is the mean collision time \cite{ZNG,NZG} associated with $C_{12}$:
\begin{eqnarray}
\frac{1}{\tau_{12}}
&\equiv&
 \frac{2 g^2}{(2\pi)^5\hbar^7} \int d{\bf p}_1
\int d{\bf p}_2 \int d{\bf p}_3 \cr
&&\times  \delta({\bf p}_1-{\bf p}_2-{\bf p}_3)
\delta(\mu_{c0}+\tilde\varepsilon_1
-\tilde\varepsilon_2-\tilde\varepsilon_3) \cr
&&\times (1+f_{10})f_{20}f_{30},
\label{tau12}
\end{eqnarray}
where $f_{i0}=[e^{\beta_0(\tilde\varepsilon_i-\mu_{c0})}-1]^{-1}$ is the static
equilibrium distribution function and $\tilde\varepsilon_i=\frac{p^2}{2m}+U_0$.

Since one can show \cite{ZNG,CJP} that $\delta\mu_{\rm diff}$
can be written as a linear combination of $\delta\tilde n$ and $\delta \tilde P$, 
the above coupled hydrodynamic equations in (\ref{lin_C}) and  (\ref{linZGN'})
are a closed set for the variables
$\delta n_c,\delta{\bf v}_c,\delta\tilde n,\delta{\bf v}_n$ and
$\delta\tilde P$.
However, it will be useful later to have an equation of motion for 
$\delta\mu_{\rm diff}$. 
This is given by [see Eq.~(86) of Ref.~\cite{ZNG}]
\begin{equation}
\frac{\partial\delta\mu_{\rm diff}}{\partial t}
=-g\bbox{\nabla}\cdot[n_{c0}(\delta{\bf v}_c-\delta{\bf v}_n)]
-\frac{1}{3}gn_{c0}\bbox{\nabla}\cdot\delta{\bf v}_n
-\frac{\delta\mu_{\rm diff}}{\tau_{\mu}}.
\label{eq_mudiff}
\end{equation}
Here $\tau_{\mu}$ is a new relaxation time governing how
$\delta\mu_{\rm diff}$ relaxes to zero, i.e., how fast diffusive equilibrium
is reached between the condensate and non-condensate components.
It is related to the collision time $\tau_{12}$ in (\ref{tau12}) by the expression
\begin{equation}
\frac{1}{\tau_{\mu}({\bf r})}=\frac{\beta g n_{c0}}{\sigma_H\tau_{12}},
\label{taumu}
\end{equation}
where the dimensionless hydrodynamic renormalization factor
$\sigma_H$ is given by
\begin{equation}
\sigma_H({\bf r})=\left[\frac{\frac{5}{2}\tilde P_0+2g\tilde n_0n_{c0}
+\frac{2}{3}g^2\tilde\gamma_0n_{c0}^2}{\frac{5}{2}\tilde P_0\tilde\gamma_0
-\frac{3}{2}g\tilde n_0^2} -1\right]^{-1},
\label{sigma_H}
\end{equation}
where $\tilde\gamma\equiv \frac{g}{k_{\rm B}T\Lambda^3}g_{1/2}(z)$.

We can now look for normal mode solutions of the linearized ZGN$'$
equations in (\ref{lin_C}) and (\ref{linZGN'}).
Assuming that these fluctuations have a
time dependence $e^{-i\omega t}$, one can solve 
(\ref{eq_mudiff}) for $\delta\mu_{\rm diff}$ to give
\begin{equation}
\delta\mu_{\rm diff}=-\frac{\tau_{\mu}}{1-i\omega\tau_{\mu}}
\left\{
g\bbox{\nabla}\cdot[n_{c0}(\delta{\bf v}_c-\delta{\bf v}_n)]
+\frac{1}{3}gn_{c0}\bbox{\nabla}\cdot\delta{\bf v}_n\right\}.
\label{eq_mudiff2}
\end{equation}
In the limit $\omega\tau_{\mu}\to 0$, one sees that $\delta\mu_{\rm diff}\to 0$.
This situation corresponds to the complete local equilibrium between the
condensate and non-condensate components, with $\tilde\mu({\bf r},t)=\mu_c({\bf r},t)$.
In this limit, one can prove that our ZGN$'$ hydrodynamics
reduces to the Landau two-fluid hydrodynamics without dissipation terms,
as discussed in detail in Ref.~\cite{ZNG}.
It is clear that fluctuation of $\delta\mu_{\rm diff}$ described by
(\ref{eq_mudiff}) [or equivalently (\ref{eq_mudiff2})] gives rise to
a new relaxational mode in addition to usual collective oscillations
of the condensate and non-condensate (for a uniform superfluid, these
are the first and second sound modes).
For a uniform Bose gas, the frequency of this new mode is given by
$\omega_{\rm R}=-i/\tau_{\mu}$ \cite{NZG}.
Thus, in general, our ZGN$'$ equations predict the existence of a new
relaxational mode, associated with the equilibration of the condensate
and non-condensate collective variables.

In Ref.~\cite{CJP}, we have extended the theory to include small deviations
from the local equilibrium distribution $\tilde f$ in (\ref{eq8}).
These give rise to new dissipative terms in the non-condensate
equations associated with the shear viscosity ($\eta$) and the thermal conductivity
($\kappa$) of the thermal cloud.
The damping of first sound, second sound and 
the relaxational mode due to the effect of normal fluid transport coefficients
was calculated in Ref.~\cite{CJP}.
In particular, we showed there that the relaxational mode was strongly coupled 
to (and renormalized by) fluctuations in the local temperature and hence the
thermal conductivity.

In summary, the ZGN$'$ hydrodynamics exhibit the physics of the coupled dynamics of the
condensate and non-condensate atoms in a clear fashion.
However the approach used in the ZGN$'$ theory has a disadvantage in that it is not
based on a small expansion parameter, in contrast to the more systematic Chapman-Enskog
procedure used here.
In Ref.~\cite{CJP}, we only included the effect of $C_{22}$ collisions
to discussing the deviation from local equilibrium.
This neglect of $C_{12}$ collisions in this connection is only justified when the
condensate density is very small compared to the non-condensate density
(since the $C_{12}$ term in (\ref{eq4}) is proportional to $n_c$).
However in a trapped gas, the $C_{12}$ collision integral is always significant
since the condensate is strongly peaked at the trap center, with a density much larger
than the non-condensate even at temperatures close to $T_{\rm BEC}$.
Thus, we must treat both $C_{12}$ and $C_{22}$ 
when considering deviations from local equilibrium.

In the following section, we present a more systematic derivation
of the two-fluid hydrodynamics, by following the standard Chapman-Enskog
procedure.
This derivation is similar to the work by Kirkpatrick
and Dorfman \cite{KD} for a {\it uniform} Bose gas.
As we discuss in Section \ref{sec:LK}, this new approach allows us to
show that the extended ZGN$'$ theory can be written in a form completely equivalent to
the Landau-Khalatnikov two-fluid hydrodynamics \cite{Khal} when we include hydrodynamic
damping.
This set of equations involves the thermal conductivity, shear viscosity and four
frequency-dependent second viscosity coefficients.
The latter are shown to arise from the fact that the condensate and non-condensate
are not in diffusive equilibrium ($\mu_c\neq\tilde\mu$), as discussed by
ZGN$'$ \cite{ZNG,NZG}.

\section{Chapman-Enskog expansion for a Bose-condensed gas}
\label{sec:CE}
\subsection{Lowest-order hydrodynamic equations}

Following the standard procedure of the Chapman-Enskog expansion
\cite{ferziger}, we introduce a small expansion parameter $\alpha$ and rewrite
the kinetic equation (\ref{eq1}) as
\begin{eqnarray}
{\partial f({\bf r},{\bf p},t) \over \partial t} + {{\bf p} \over m} 
\cdot \bbox{\nabla}_{{\bf r}} f({\bf r},{\bf p},t) &-& \bbox{\nabla}_{{\bf r}} U \cdot
\bbox{\nabla}_{{\bf p}} f({\bf r} ,{\bf p},t) \cr
&=& \frac{1}{\alpha}\left(C_{12}[f] + C_{22}[f]\right).
\label{kineq2}
\end{eqnarray}
This expansion parameter $\alpha$ will be eventually taken to be 1,
but allows one to develop a perturbative solution of (\ref{kineq2}).
In order to solve the quantum kinetic equation, 
we formally expand the distribution function $f({\bf r},{\bf p},t)$ in powers of $\alpha$:
\begin{equation}
f=f^{(0)}+\alpha f^{(1)}+ \cdots.
\label{f_alpha}
\end{equation}
Using this expansion (\ref{f_alpha}), we can also expand the various hydrodynamic variables
in (\ref{hydro_general})
\begin{eqnarray}
\tilde n&=&\tilde n^{(0)}+\alpha\tilde n^{(1)}+\cdots, ~~
P_{\mu\nu}=P_{\mu\nu}^{(0)}+\alpha P_{\mu\nu}^{(1)}+\cdots, ~~
{\bf Q}={\bf Q}^{(0)}+\alpha{\bf Q}^{(1)}+\cdots, \cr
\tilde\epsilon&=&\tilde\epsilon^{(0)}+\alpha\tilde\epsilon^{(1)}+\cdots.
\label{expand}
\end{eqnarray}
The superscript $(0)$ denotes the local equilibrium solution (see below) which is
determined by the collision integrals (formally when $\alpha\to 0$).
We also redefine the source function $\Gamma_{12}$ in (\ref{def_G12}) as
\begin{equation}
\Gamma_{12}\equiv\frac{1}{\alpha}\int\frac{d{\bf p}}{(2\pi\hbar)^3}
C_{12}[f,\Phi]=\frac{1}{\alpha}
(\Gamma_{12}^{(0)}+\alpha\Gamma_{12}^{(1)}+\alpha^2\Gamma_{12}\cdots).
\label{redef_G12}
\end{equation}
We also have an expansion for the condensate wavefunction
\begin{equation}
\Phi=\Phi^{(0)}+\alpha \Phi^{(1)}+\cdots.
\end{equation}

In this expansion, however, we assume that the total local density $n(\equiv n_c+\tilde n)$
is not altered by the higher order correction terms $f^{(i)}~(i\geq 1)$ in (\ref{f_alpha}).
That is, we have
\begin{equation}
n_c=n_c^{(0)}+\alpha n_c^{(1)}+\cdots,
\end{equation}
but
\begin{equation}
n=n_c^{(0)}+\tilde n^{(0)}, ~n_c^{(i)}+\tilde n^{(i)}=0 ~(i\geq 1).
\end{equation}
We also assume that non-local correction terms $f^{(i)}$ make
no contribution to the non-condensate velocity fields ${\bf v}_n$ or to the phase
$\theta$ of the condensate wavefunction (and hence to the condensate velocity ${\bf v}_c$).
Finally, the condensate chemical potential in (\ref{mu_c})
(we work within the Thomas-Fermi approximation) is given in the expansion
\begin{equation}
\mu_c({\bf r},t)=U_{\rm ext}({\bf r})+g[n({\bf r},t)+\tilde n({\bf r},t)]
=\mu_c^{(0)}({\bf r},t)+\alpha\mu_c^{(1)}({\bf r},t)+\cdots,
\end{equation}
with 
\begin{equation}
\mu_c^{(0)}\equiv U_{\rm ext}+g(n+\tilde n^{(0)}),~\mu_c^{(1)}=g\tilde n^{(1)}.
\label{muc01}
\end{equation}

Using the expansion (\ref{f_alpha}) in the kinetic equation (\ref{kineq2}), 
we find that the lowest order solution $f^{(0)}$ is determined from
\begin{equation}
C_{12}[f^{(0)},\Phi^{(0)}]+C_{22}[f^{(0)}]=0.
\label{eqforf0}
\end{equation}
The unique solution of (\ref{eqforf0}) is given by the ``diffusive local equilibrium''
Bose distribution function, namely
\begin{equation}
f^{(0)}({\bf r},{\bf p},t)=
\frac{1}{e^{\beta({\bf r},t)
[\frac{1}{2m}({\bf p}-m{\bf v}_n({\bf r},t))^2+U({\bf r},t)-
\tilde \mu^{(0)}({\bf r},t)]}-1}
\, .
\label{f0}
\end{equation}
Here the local equilibrium non-condensate chemical potential
$\tilde\mu^{(0)}$ is given by the condition that $C_{12}[f^{(0)},\Phi^{(0)}]=0$,
which gives
\begin{equation}
\tilde \mu^{(0)}=\mu_c^{(0)}+\frac{m}{2}({\bf v}_n-{\bf v}_c)^2. 
\label{muc0}
\end{equation}
Using (\ref{muc01}), this is equivalent to
\begin{equation}
\tilde \mu^{(0)}=\mu_c^{(0)}+\frac{m}{2}({\bf v}_n-{\bf v}_c)^2 
=U_{\rm ext}+gn+g\tilde n^{(0)}+\frac{m}{2}({\bf v}_n-{\bf v}_c)^2 ,
\label{muc02}
\end{equation}
in conjunction with
\begin{equation}
\tilde n^{(0)}({\bf r},t) = \int{d{\bf p}\over (2\pi\hbar)^3}
f^{(0)}({\bf r, p}, t)
= {1\over\Lambda^3} g_{3/2}(z^{(0)})\ .
\label{ntilde0}
\end{equation}
Here $z^{(0)}({\bf r},t)\equiv e^{\beta[\tilde\mu^{(0)}({\bf r},t)-U({\bf r},t)]}$ is the local
fugacity in diffusive local equilibrium.

It is important to appreciate that diffusive equilibrium
is not defined by the distribution function $f^{(0)}$ alone, but is determined
self-consistently with the non-condensate chemical potential as given by
(\ref{muc0}).
One may immediately verify that $f^{(0)}$ satisfies $C_{22}[f^{(0)}]=0$,
independent of the value of $\tilde \mu^{(0)}$.  
In contrast, $C_{12}[f^{(0)},\Phi^{(0)}]=0$
{\it only} if the local chemical potential of the thermal cloud is given by (\ref{muc0})
and the condensate and non-condensate densities are determined self-consistently.
Of course, it immediately follows that since $C_{12}[f^{(0)},\Phi^{(0)}]=0$,
we have $\Gamma_{12}^{(0)}=\Gamma_{12}[f^{(0)},\Phi^{(0)}]=0$ and hence
(\ref{redef_G12}) reduces to 
\begin{equation}
\Gamma_{12}=\Gamma_{12}^{(1)}+\alpha \Gamma_{12}^{(2)}
+\cdots.
\label{G12_alpha}
\end{equation}

Using the local distribution function (\ref{f0}) 
to evaluate the moments in (\ref{eq35b}), we find that 
the heat current ${\bf Q}^{(0)}({\bf r},t) = 0$, and 
\begin{equation}
P_{\mu\nu}^{(0)} ({\bf r}, t) = \delta_{\mu\nu}{\tilde P}^{(0)}({\bf r}, t)
\equiv\delta_{\mu\nu}\int{d{\bf p}\over (2\pi)^3} {({\bf p}-m{\bf v}_n)^2\over 3m}
f^{(0)}({\bf r, p}, t)
=\delta_{\mu\nu}\frac{1}{\beta\Lambda^3}g_{5/2}(z^{(0)}).
\label{ptilde0}
\end{equation}
The local kinetic energy density  is given by
$\tilde \epsilon^{(0)}({\bf r},t) ={3\over 2} \tilde P^{(0)} ({\bf r},t)$.

To summarize, the lowest-order hydrodynamic equations for the non-condensate are
given by
\begin{mathletters}
\begin{eqnarray}
&&{\partial{\tilde n}\over\partial t}+\bbox{\nabla}\cdot 
(\tilde n{\bf v}_n) = \Gamma_{12}^{(1)}
\label{eq_ntilde0}
\,, \\
&&m{\tilde n}\left({\partial\over\partial t}+{\bf v}_n\cdot 
\bbox{\nabla}\right) {\bf v}_n=-\bbox{\nabla} \tilde P
-{\tilde n}\bbox{\nabla} U
-m({\bf v}_n-{\bf v}_c)\Gamma_{12}^{(1)}\,, 
\label{eq_vn0}\\
&&{\partial\tilde P\over\partial t} +
\nabla\cdot(\tilde P{\bf v}_n) = -{2\over 3}\tilde P \bbox{\nabla}
\cdot{\bf v}_n + {2\over 3} \left[\frac{1}{2}m({\bf v}_n-{\bf v}_c)^2
+\mu_c-U\right]\Gamma_{12}^{(1)},
\label{eq_ptilde0}
\end{eqnarray}
\label{lowest-eqs}
\end{mathletters}

\noindent
where $\tilde n=\tilde n^{(0)}$, $\tilde P=\tilde P^{(0)}$ and $\mu_c=\mu_c^{(0)}$
are given by (\ref{muc0}), (\ref{ntilde0}) and (\ref{ptilde0}).
It should be noted that the above equations involve the source term $\Gamma_{12}^{(1)}$.
Even though $C_{12}[f^{(0)}]=0$,
one sees from (\ref{G12_alpha}) that the lowest order contribution
is in fact given by $\Gamma_{12}^{(1)}$, which
involves the contribution from the next order correction $f^{(1)}$.
Later we will derive an explicit expression for $\Gamma_{12}$ when we include the 
effect of deviations from the local equilibrium distribution and transport
processes.
Here we only display the result for the lowest order contribution which enter
into (\ref{lowest-eqs}) [see also (\ref{eq_G12})]:
\begin{equation}
\Gamma_{12}^{(1)}({\bf r},t)=\sigma_H\left\{
\bbox{\nabla}\cdot[n_c({\bf v}_c-{\bf v}_n)]
+\frac{1}{3}n_c\bbox{\nabla}\cdot{\bf v}_n\right\},
\label{gamma1}
\end{equation}
where $\sigma_H$ is defined by (\ref{sigma_H}).

It is important to note that even though $\Gamma_{12}^{(1)}$ involves an integral
over the collision integral $C_{12}$ (see (\ref{redef_G12})), the expression in (\ref{gamma1})
does not involve any collision time.
The expression for $\Gamma_{12}^{(1)}$ in (\ref{gamma1}) is consistent with
the ZGN$'$ result for $\delta\mu_{\rm diff}$ given in (\ref{eq_mudiff2}) 
in the limit $\omega\tau_{\mu}\to 0$ [using (\ref{G12}) and (\ref{sigma_H})].
We recall that in this limit, one has $\delta\mu_{\rm diff}\to 0$ and
thus $\tilde f$ in (\ref{eq8}) reduces to $f^{(0)}$ in (\ref{f0}).
Therefore the hydrodynamic equations given in (\ref{lowest-eqs}) are 
equivalent, in the $\omega\tau_{\mu}\to 0$ limit,
to those given by the ZGN$'$ theory.
As noted in Ref.~\cite{ZNG},
these coupled lowest-order hydrodynamic equations
in (\ref{lowest-eqs}) and (\ref{hydro-C}) can be combined 
and also shown to be precisely equivalent to the Landau two-fluid equations
without dissipation due to the transport processes.
In Section \ref{sec:LK}, we prove this equivalence in the more general case
when dissipation is included.

\subsection{Two-fluid equations with hydrodynamic dissipation}

We next consider the deviation (\ref{f_alpha}) from the local equilibrium
distribution function $f^{(0)}$ to first order in the Chapman-Enskog expansion.
This deviation $f^{(0)}$ gives rise to additional dissipative
terms in the hydrodynamic equations.
As usual, in determining the dissipative terms, we restrict ourselves to
terms of first order in the velocity fields ${\bf v}_n$ and ${\bf v}_c$.
Following Refs.~\cite{KD,NG,CJP}, we write the first correction term in
(\ref{f_alpha}) in the form
\begin{equation}
f^{(1)}=f^{(0)}({\bf r},{\bf p},t) [1+f^{(0)}({\bf r},{\bf p},t)]
\psi({\bf r},{\bf p},t),
\label{correction}
\end{equation}
and work with $\psi({\bf r},{\bf p},t)$.
To first order in $\alpha$, the $C_{22}$ and $C_{12}$ collision terms in (\ref{eq2})
reduce to ($f=f^{(0)}+\alpha f^{(1)}$).
\begin{eqnarray}
\frac{1}{\alpha}
C_{22}[f]&\simeq&C_{22}[f^{(1)}]\simeq \frac{2g^2}{(2\pi)^5\hbar^7}\int d{\bf p}_2
\int d{\bf p}_3 \int d{\bf p}_4 \cr
&&\times \delta({\bf p}+{\bf p}_2-{\bf p}_3-{\bf p}_4)\delta(\tilde\varepsilon_{p_1}
+\tilde\varepsilon_{p_2}-\tilde\varepsilon_{p_3}-\tilde\varepsilon_{p_4}) \cr
&& \times f^{(0)} f^{(0)}_2\left(1+f^{(0)}_3\right)
\left(1+f^{(0)}_4\right)
(\psi_3+\psi_4-\psi_2-\psi) \cr
&& \equiv \hat L_{22}[\psi].
\label{L22}
\end{eqnarray}
\begin{eqnarray}
\frac{1}{\alpha}C_{12}[f,\Phi]&\simeq&
 \frac{2 g^2 n_c}{(2\pi)^2\hbar^4} \int d{\bf p}_1
\int d{\bf p}_2 \int d{\bf p}_3 \cr
&&\times  \delta(m{\bf v}_c+{\bf p}_1-{\bf p}_2-{\bf p}_3)
\delta(\varepsilon_c^{(0)}+\tilde\varepsilon_{p_1}
-\tilde\varepsilon_{p_2}-\tilde\varepsilon_{p_3}) \cr
&&\times  [\delta({\bf p}-{\bf p}_1)-\delta({\bf p}-{\bf p}_2)
-\delta({\bf p}-{\bf p}_3)] \cr
&&\times (1+f^{(0)}_1)f^{(0)}_2f^{(0)}_3
(-\beta\mu_c^{(1)}+\psi_2+\psi_3-\psi_1) \cr 
&& \equiv -\beta g\tilde n^{(1)}\hat L_{12}[1]+\hat L_{12}[\psi],
\label{C12_1}
\end{eqnarray}
where $\varepsilon_c^{(0)}=\mu_c^{(0)}+\frac{1}{2}mv_c^2$ and $\mu_c^{(0)}$
is given by (\ref{muc01}).
The linearized $\hat L_{12}$ operator is defined by
\begin{eqnarray}
\hat L_{12}[\psi]
&\equiv&
 \frac{2 g^2 n_c}{(2\pi)^2\hbar^4} \int d{\bf p}_1
\int d{\bf p}_2 \int d{\bf p}_3 \cr
&&\times  \delta(m{\bf v}_c+{\bf p}_1-{\bf p}_2-{\bf p}_3)
\delta(\varepsilon_c^{(0)}+\tilde\varepsilon_{p_1}
-\tilde\varepsilon_{p_2}-\tilde\varepsilon_{p_3}) \cr
&&\times  [\delta({\bf p}-{\bf p}_1)-\delta({\bf p}-{\bf p}_2)
-\delta({\bf p}-{\bf p}_3)] \cr
&&\times (1+f^{(0)}_1)f^{(0)}_2f^{(0)}_3
(\psi_2+\psi_3-\psi_1).
\label{L12}
\end{eqnarray}

Using (\ref{L22})-(\ref{L12}) and expanding the kinetic equation 
(\ref{kineq2}) to first order in $\alpha$, we find that 
the first non-local correction $f^{(1)}$ is determined by the equation
\begin{eqnarray}
&&{\partial^0 f^{(0)}({\bf r},{\bf p},t) \over \partial t} + {{\bf p} \over m} 
\cdot \bbox{\nabla}_{{\bf r}} f^{(0)}({\bf r},{\bf p},t) - \bbox{\nabla}_{{\bf r}} U \cdot
\bbox{\nabla}_{{\bf p}} f^{(0)}({\bf r} ,{\bf p},t) \cr
&&= -\beta g\tilde n^{(1)}\hat L_{12}[1]+\hat L_{12}[\psi] + \hat L_{22}[\psi].
\label{eq_f1}
\end{eqnarray}
Here $\partial^0/\partial t$ means that we use the lowest order
hydrodynamic equations given by (\ref{lowest-eqs})
in evaluating time derivatives of 
${\bf v}_n$, $\tilde\mu$, $T$ and $U$. 
The resulting linearized equation which determines the function $\psi$ is 
(for details, see Appendix A)
\begin{eqnarray}
&&\left\{\frac{{\bf u}\cdot\bbox{\nabla}T}{T}\left[\frac{mu^2}{2k_{\rm B}T}-
\frac{5g_{5/2}(z)}{2g_{3/2}(z)}\right]+\frac{m}{k_{\rm B}T}D_{\mu\nu}
\left(u_{\mu}u_{\nu}-\frac{1}{3}\delta_{\mu\nu}u^2\right) \right. \cr
&&
+\left.\left(\sigma_2+\frac{mu^2}{3k_{\rm B}T}\sigma_1
 \right)\frac{\Gamma_{12}^{(1)}}{\tilde n^{(0)}}\right\}
f^{(0)}(1+ f^{(0)})+\beta g\tilde n^{(1)}\hat L_{12}[1]
=\hat L_{12}[\psi]+\hat L_{22}[\psi]\equiv \hat L[\psi],
\label{linpsi}
\end{eqnarray}
where the thermal velocity ${\bf u}$ is defined by 
$m{\bf u}\equiv {\bf p}-m{\bf v}_n$ and $z=z^{(0)}$.
The dimensionless thermodynamic functions $\sigma_1$, $\sigma_2$ in (\ref{linpsi}) are 
defined by
\begin{eqnarray}
\sigma_1({\bf r},t)&\equiv&\frac{\gamma^{(0)}\tilde n^{(0)}
\left[\tilde \mu^{(0)}-U\right]-\frac{3}{2}[\tilde n^{(0)}]^2}
{\frac{5}{2}\tilde P^{(0)}\gamma^{(0)}-\frac{3}{2}[\tilde n^{(0)}]^2}, \cr
\sigma_2({\bf r},t)&\equiv&\beta\frac{\frac{5}{2}\tilde P^{(0)}\tilde n^{(0)}
-[\tilde n^{(0)}]^2 \left[\tilde \mu^{(0)}-U\right]}
{\frac{5}{2}\tilde P^{(0)}\gamma^{(0)}-\frac{3}{2}[\tilde n^{(0)}]^2},
\label{sigma12}
\end{eqnarray}
where $\gamma^{(0)}({\bf r},t)\equiv \frac{\beta}{\Lambda^3}g_{1/2}(z^{(0)}({\bf r},t))
=\tilde\gamma^{(0)}/g$.
We note that $C_{12}$ enters in three separate places in (\ref{linpsi}).

The linearized collision operators $\hat L_{12}$ and $\hat L_{22}$
satisfy the conditions
\begin{eqnarray}
&&\hat L_{12}[{\bf p}-m{\bf v}_c]=0, 
~~\hat L_{12}[\tilde\varepsilon_p-\varepsilon_c^{(0)}]=0, \cr
&&\hat L_{22}[1]=0, ~~
\hat L_{22}[{\bf p}]=0, ~~\hat L_{22}[\tilde\varepsilon_p]=0.
\end{eqnarray}
In order to have a unique solution of (\ref{linpsi}) for $\psi$, we impose
the following additional constraints:
\begin{mathletters}
\begin{eqnarray}
&&\int d{\bf p} ~{\bf u} \ f^{(0)}(1+f^{(0)})\psi=0,  \label{const_vn}\\
&&\int d{\bf p}\left(\frac{m}{2}u^2+U-\tilde\mu^{(0)}\right)
f^{(0)}(1+f^{(0)})\psi \cr
&&~~~~=\frac{1}{\beta}\int d{\bf p} \ln (1+f^{(0)^{-1} })f^{(1)}=0.
\label{const_s}
\end{eqnarray}
\label{const}
\end{mathletters}
\noindent

\noindent
Physically, the constraint (\ref{const_vn}) means that the deviation from local
equilibrium make no contribution to the local velocity field ${\bf v}_n$
defined in (\ref{vn}).
As we discuss in more detail in Section \ref{sec:LK}, the constraint (\ref{const_s}) means
that the total energy density and the local entropy density are not altered by
the deviation $f^{(1)}$.
They have the the same value as given by $f^{(0)}$.

Since (\ref{linpsi}) is a linear integral equation for $\psi$, one may write the most general
solution in the following form \cite{ferziger}:
\begin{equation}
\psi({\bf r},{\bf p},t)=\frac{\bbox{\nabla}T\cdot{\bf u}}{T}A(u)+D_{\mu\nu}
\left(u_{\mu}u_{\nu}-\frac{1}{3}u^2\delta_{\mu\nu}\right)B(u)
+\Gamma_{12}^{(1)}D(u),
\label{solution}
\end{equation}
where the dependence on $({\bf r},t)$ is left implicit and $u_{\mu}$ is a
component of the thermal velocity.
Here the functions $A(u),B(u)$ and $D(u)$ are given by the solution to the
following three linearized integral equations:
\begin{mathletters}
\begin{eqnarray}
&&{\bf u}\left[\frac{mu^2}{2k_{\rm B}T}-
\frac{5g_{5/2}(z)}{2g_{3/2}(z)}\right]f^{(0)}(1+\tilde f^{(0)})=
\hat L[{\bf u}A(u)], \label{eqforA}\\
&&
\frac{m}{k_{\rm B}T}
\left(u_{\mu}u_{\nu}-\frac{1}{3}\delta_{\mu\nu}u^2\right)
f^{(0)}(1+f^{(0)})
=\hat L[\left(u_{\mu}u_{\nu}-\frac{1}{3}\delta_{\mu\nu}u^2\right)B(u)], 
\label{eqforB} \\
&&\left(\sigma_2+\frac{mu^2}{3k_{\rm B}T}\sigma_1\right)\frac{1}{\tilde n^{(0)}}f^{(0)}(1+f^{(0)})
+\frac{\beta g\tilde n^{(1)}}{\Gamma_{12}^{(1)}} \hat L_{12}[1]=\hat L[D(u)].
\label{eqforD}
\end{eqnarray}
\label{eqforABD}
\end{mathletters}
For the constraints (\ref{const}) to be satisfied, we also need to require that
\begin{mathletters}
\begin{eqnarray}
&&\int \frac{d{\bf p}}{(2\pi\hbar)^2} f^{(0)}(1+f^{(0)})u^2 A(u)=0, 
\label{constA}
\\
&&\int \frac{d{\bf p}}{(2\pi\hbar)^2} f^{(0)}(1+f^{(0)})
\left(\frac{mu^2}{2}+U-\tilde \mu^{(0)}\right) D(u)=0.
\label{constD}
\end{eqnarray}
\end{mathletters}

Using the solution for $\psi$ given in (\ref{solution}),
one finds that the corrections due to $f^{(1)}$ in (\ref{correction})
to the various hydrodynamic variables are given by
\begin{eqnarray}
\tilde n^{(1)}&=&\int \frac{d{\bf p}}{(2\pi\hbar)^3}
f^{(0)}(1+f^{(0)})D(u)\Gamma_{12}^{(1)}({\bf r},t)\equiv -\tau\Gamma_{12}^{(1)},  
\label{n1} \\
P^{(1)}_{\mu\nu}&=&\delta_{\mu\nu}\tilde P^{(1)}
-2\eta\left[D_{\mu\nu}-\frac{1}{3}
{\rm Tr}D \delta_{\mu\nu}\right], \\
{\bf Q}^{(1)}&=&-\kappa\bbox{\nabla}T,
\end{eqnarray}
with
\begin{eqnarray}
\tilde P^{(1)}&=&\tau\frac{2}{3}(U-\tilde \mu^{(0)})\Gamma_{12}^{(1)}
\simeq \tau\frac{2}{3}gn_c^{(0)}\Gamma_{12}^{(1)}, \\
\tilde\epsilon^{(1)}&=&\frac{3}{2}\tilde P^{(1)}.
\end{eqnarray}
We note that $\tilde n$ and $\tilde P$ are both altered by an amount proportional
to $\Gamma_{12}^{(1)}$.
The transport coefficients $\eta$ and $\kappa$ are associated with the functions
$A(u)$ and $B(u)$,
\begin{mathletters}
\begin{eqnarray}
\eta&=&-\frac{m}{15}\int\frac{d{\bf p}}{(2\pi\hbar)^3}
u^4B(u)f^{(0)}(1+f^{(0)}), 
\label{eta}
\\
\kappa&=&-\frac{m}{6T}\int\frac{d{\bf p}}{(2\pi\hbar)^3}
u^4A(u)f^{(0)}(1+f^{(0)}).
\label{kappa}
\end{eqnarray}
\label{eta-kappa}
\end{mathletters}

The relaxation time $\tau$ defined in (\ref{n1}), namely
\begin{equation}
\tau=-\int\frac{d{\bf p}}{(2\pi\hbar)^3}f^{(0)}(1+f^{(0)})D(u),
\label{tau}
\end{equation}
plays a crucial role in the subsequent analysis.
Using (\ref{n1}) in (\ref{eqforD}), one can rewrite the integral equation for $D(u)$
in the form
\begin{equation}
\left(\sigma_2+\frac{mu^2}{3k_{\rm B}T}\sigma_1\right)\frac{1}{\tilde n^{(0)}}f^{(0)}(1+f^{(0)})
-\tau\beta g \hat L_{12}[1]=\hat L[D(u)].
\end{equation}
In Section \ref{sec:transport}, we solve the three linearized equations listed in
(\ref{eqforABD}).
It will be shown there that the solution for the function $D(u)$ is
\begin{equation}
D(u)=-\frac{\tau_{\mu}}{\tilde n^{(0)}}\left(\sigma_2+\frac{mu^2}{3k_{\rm B}T}
\sigma_1\right).
\label{Du}
\end{equation}
Using this, one finds that $\tau$ can be identified with the relaxation time $\tau_{\mu}$
defined in (\ref{taumu}).
In the present discussion, the physical meaning of the relaxation time $\tau_{\mu}$
can be clearly seen by writing the source function $\Gamma_{12}^{(1)}$ 
in the form (see (\ref{n1}) and (\ref{expand})) 
\begin{equation} 
\Gamma_{12}^{(1)}=-\frac{\tilde n^{(1)}}{\tau}=
-\frac{\tilde n-\tilde n^{(0)}}{\tau_{\mu}}.
\end{equation}
This kind of relaxation term in the two-fluid hydrodynamic equations such as
(\ref{eq_nc}) and (\ref{eq_ntilde0})
was also discussed in a pioneering paper by Miyake and Yamada~\cite{MY}
in discussing the liquid $^4$He near the superfluid transition
(where a phenomenological relaxation time equivalent to $\tau_{\mu}$ was introduced).

In summary, we have obtained the following hydrodynamic equations for 
the non-condensate including the normal fluid transport coefficients
(we now set the expansion parameter $\alpha=1$):
\begin{mathletters}
\label{eqhydro}
\begin{eqnarray}&&\frac{\partial \tilde n}{\partial t}
+\bbox{\nabla}\cdot(\tilde n{\bf v}_n)=\Gamma_{12}
\label{eq_tilden}\\
&&m\tilde n\left(\frac{\partial}{\partial t}+{\bf v}_n\cdot \bbox{\nabla}\right)v_{n\mu}
+\frac{\partial \tilde P}{\partial x_{\mu}}+\tilde n
\frac{\partial U}{\partial
x_{\mu}}=-m(v_{n\mu}-v_{c\mu})\Gamma_{12} \cr
&& \ \ \ \ \ \ \ \ \ \ \ \ \ \ \ \ \ \ \ \ \ \ \ \ 
+\frac{\partial}{\partial x_{\nu}}\left\{2\eta\left[D_{\mu\nu}-\frac{1}{3}
({\rm Tr}D)\delta_{\mu\nu} \right]\right\},
\label{eq_vn}\\
&&\frac{\partial \tilde \epsilon}{\partial t}+\bbox{\nabla}
\cdot(\tilde\epsilon{\bf v}_n)
+(\bbox{\nabla}\cdot{\bf v}_n)\tilde P=\left[\frac{1}{2}m({\bf v}_n-{\bf v}_c)^2
+\mu_c-U\right]\Gamma_{12} \cr
&& \ \ \ \ \ \ \ \ \ \ \ \ \ \ \ \ \ \ \ \ \ \ \ \ \ \ 
+\bbox{\nabla}\cdot(\kappa\bbox{\nabla} T)
+2\eta\left[ D_{\mu\nu}-\frac{1}{3}({\rm Tr}D)\delta_{\mu\nu}\right]^2,
\label{eq26c}
\end{eqnarray}
\label{hydro-n}
\end{mathletters}
where $\tilde n$ and $\tilde P$ are given by
\begin{eqnarray}
&&\tilde n=\tilde n^{(0)}-\tau_{\mu}\Gamma_{12}^{(1)},  \label{ntildetot}\\
&&\tilde P=\tilde P^{(0)}+\tau_{\mu}\frac{2}{3}gn_c^{(0)}\Gamma_{12}^{(1)}, \label{ptildetot}
\end{eqnarray}
and $\tilde\epsilon=\frac{3}{2}\tilde P$.
Here $\tilde n^{(0)}$ and $\tilde P^{(0)}$ are given by (\ref{ntilde0}) and (\ref{ptilde0}),
respectively.
The equivalent ``quantum"
hydrodynamic equations for the condensate are given in (\ref{hydro-C}),
where the condensate chemical potential is given by
\begin{equation}
\mu_c=\mu_c^{(0)}-g\tau_{\mu}\Gamma_{12}^{(1)}.
\label{muctot}
\end{equation}

We now derive an equation for the function $\Gamma_{12}$.
Using (\ref{ntildetot}) and (\ref{eq_tilden}), one obtains
\begin{equation}
\frac{\partial \tilde n}{\partial t}=
\frac{\partial \tilde n^{(0)}}{\partial t}-\tau_{\mu}\frac{\partial \Gamma_{12}}{\partial t}
=-\bbox{\nabla}\cdot(\tilde n{\bf v}_n)+\Gamma_{12}.
\label{eq_totn}
\end{equation}
Using the explicit expression for $\tilde n^{(0)}$ given in (\ref{ntilde0}),
one obtains
\begin{equation}
\frac{\partial \tilde n^{(0)}}{\partial t}
=\left(\frac{3}{2}\tilde n+g \gamma n_c\right)\frac{1}{T}\frac{\partial T}{\partial t}
+g\gamma\left(\frac{\partial \tilde n^{(0)}}{\partial t}-
\frac{\partial n}{\partial t}\right).
\label{eq70}
\end{equation}
Using the continuity equation for the total density $n$, one finds that
(\ref{eq70}) reduces to
\begin{equation}
\frac{\partial \tilde n^{(0)}}{\partial t}
=\frac{1}{1-g\gamma}
\left[\left(\frac{3}{2}\tilde n +g\gamma n_c\right)\frac{1}{T}
\frac{\partial T}{\partial t}+\bbox{\nabla}\cdot(\tilde n{\bf v}_n
+n_c{\bf v}_c) \right].
\label{eq_n0}
\end{equation}
Using (\ref{eq_n0}) in (\ref{eq_totn}), one finds
\begin{equation}
\tau_{\mu}\frac{\partial \Gamma_{12}}{\partial t}+\Gamma_{12}
=\frac{1}{1-g\gamma}
\left[\left(\frac{3}{2}\tilde n +g\gamma n_c\right)\frac{1}{T}
\frac{\partial T}{\partial t}+\bbox{\nabla}\cdot(\tilde n{\bf v}_n
+n_c{\bf v}_c) \right]
+\bbox{\nabla}\cdot(\tilde n{\bf v}_n).
\label{G12_1}
\end{equation} 
We next use (\ref{ptildetot}) in the equation for $\tilde P$
(given by (\ref{eq26c})):
\begin{equation}
\frac{\partial \tilde P}{\partial t}
=\frac{\partial \tilde P^{(0)}}{\partial t}
+\tau_{\mu}\frac{2}{3}gn_c\frac{\partial\Gamma_{12}}{\partial t}
=-\bbox{\nabla}\tilde P\cdot{\bf v}_n-\frac{5}{3}\tilde P \bbox{\nabla}\cdot{\bf v}_n
-\frac{2}{3}gn_c\Gamma_{12}+\frac{2}{3}\bbox{\nabla}\cdot(\kappa\bbox{\nabla}T).
\label{eq73}
\end{equation}
Using the expression for $\tilde P^{(0)}$ given in (\ref{ptilde0}), one finds
\begin{equation}
\frac{\partial \tilde P^{(0)}}{\partial t}=
\left(\frac{5}{2}\tilde P+g\tilde n n_c\right)\frac{1}{T}\frac{\partial T}{\partial t}
+g\tilde n\frac{\partial \tilde n^{(0)}}{\partial t}
+g\tilde n \bbox{\nabla}\cdot(\tilde n{\bf v}_n+n_c{\bf v}_c).
\end{equation}
Substituting this into (\ref{eq73}) in conjunction with (\ref{eq_totn}), we obtain
\begin{eqnarray}
-\left(\frac{2}{3}gn_c+g\tilde n\right)\left(\tau_{\mu}\frac{\partial \Gamma_{12}}
{\partial t}+\Gamma_{12}\right)
&=&\left(\frac{5}{2}\tilde P+g\tilde n n_c\right)\frac{1}{T}\frac{\partial T}{\partial t}
+g\tilde n\bbox{\nabla}\cdot(n_c{\bf v}_c)+\bbox{\nabla}\tilde P\cdot{\bf v}_n  \cr
&&+\frac{5}{3}\tilde P\bbox{\nabla}\cdot{\bf v}_n
-\frac{2}{3}\bbox{\nabla}\cdot(\kappa\bbox{\nabla}T).
\label{G12_2}
\end{eqnarray}
One may now combine (\ref{G12_1}) and (\ref{G12_2})
to eliminate $\partial T/\partial t$ from these two equations.
After a certain amount of rearrangement, we finally obtain our desired equation of
motion for $\Gamma_{12}$:
\begin{equation}
\tau_{\mu}\frac{\partial \Gamma_{12}}{\partial t}+\Gamma_{12}
=\sigma_H\left \{ \bbox{\nabla}\cdot[n_c({\bf v}_c-{\bf v}_n]
+\frac{1}{3}n_c\bbox{\nabla}\cdot{\bf v}_n \right \}
-\frac{2}{3}\frac{\sigma_H\sigma_1}{g}\bbox{\nabla}\cdot
(\kappa\bbox{\nabla}T).
\label{eq_G12}
\end{equation}
If we keep the expansion parameter $\alpha$ and expand $\Gamma_{12}$ as in
(\ref{G12_alpha}), namely 
$\Gamma_{12}=\Gamma_{12}^{(1)}+\alpha\Gamma_{12}^{(2)}$ (we recall that $\Gamma_{12}^{(0)}=0$),
we find $\Gamma_{12}^{(1)}$ is given by (\ref{gamma1}) and 
\begin{equation}
\Gamma_{12}^{(2)}
=-\tau_{\mu}\frac{\partial}{\partial t}\Gamma_{12}^{(1)}
-\frac{2}{3}\frac{\sigma_H\sigma_1}{g\tilde n}\bbox{\nabla}\cdot
(\kappa\bbox{\nabla}T).
\end{equation}

In closing this section, we discuss the relation between
the analysis given in this Section and the ZGN$'$ theory~\cite{ZNG}
reviewed in Section~\ref{zgn'}.
In this Section, we started with the complete local equilibrium
distribution given by (\ref{f0}). We then included the deviation from local equilibrium,
as given by (\ref{correction}) with (\ref{solution}).
We showed that the deviation from $f^{(0)}$ associated with $D(u)$ in (\ref{solution})
gives rise to the corrections to the local thermodynamic quantities $\tilde n$, $\tilde P$
and $\tilde\epsilon$.
Such corrections did not arise when we
included the deviation from $\tilde f$ in the ZGN$'$ hydrodynamics~\cite{CJP}.
However, one can show that the type of contribution associated with $D(u)$
is, in fact, already contained in the lowest-order ZGN$'$ distribution function $\tilde f$
given by (\ref{eq8}).
To see this, it is convenient to linearize the distribution function
around {\it static} equilibrium, using $f\simeq f_0+\delta f$.
In the ZGN$'$ theory \cite{ZNG}, one can show that
\begin{equation}
\delta f=\beta_0f_0(1+f_0)\left[\frac{\delta \tilde T}{T_0}
\left(\frac{p^2}{2m}+U_0-\mu_{c0}\right)+{\bf p}\cdot{\bf v}_n-2g\delta n
+\delta \tilde \mu\right].
\label{df_zgn}
\end{equation}
Here we have denoted the temperature fluctuation as $\delta \tilde T$ to make
a distinction from the temperature defined in the diffusive local equilibrium
distribution function (\ref{f0})(we will find that $\delta\tilde T\neq\delta T$).
In the present theory, in contrast, one finds (ignoring the terms in (\ref{solution}) associated 
with the functions $A$ and $B$)
\begin{eqnarray}
\delta f&=&\beta_0f_0(1+f_0)\left[\frac{\delta T}{T_0}
\left(\frac{p^2}{2m}+U_0-\mu_{c0}\right)+{\bf p}\cdot{\bf v}_n-2g\delta n 
+\delta \mu_c^{(0)}\right] \cr ~~ && \cr
&&+f_0(1+f_0)D(u)\delta \Gamma_{12}.
\label{df_ng}
\end{eqnarray}

The first term in (\ref{df_ng}) represents the deviation from $f_0$
included in $f^{(0)}$ while the second term is due to $f^{(1)}$.
Using the explicit solution for $D(u)$ given by (\ref{Du})
(derived in Section \ref{sec:transport}), we find that (\ref{df_ng}) can
be written as
\begin{eqnarray}
\delta f&=&\beta_0f_0(1+f_0)\Bigg[\left(\frac{\delta T}{T_0}-\frac{2\sigma_1\tau_{\mu}}
{3\tilde n_0}\delta\Gamma_{12}\right)
\left(\frac{p^2}{2m}+U_0-\mu_{c0}\right)+{\bf p}\cdot{\bf v}_n-2g\delta n  \cr
&&+\delta \mu_c^{(0)}-g(\sigma_H^{-1}+1)\tau_{\mu}\delta\Gamma_{12}\Bigg].
\label{df_ng2}
\end{eqnarray}
We note that this linearized distribution function has the
same form as the ZGN$'$ distribution function in (\ref{df_zgn}), but with a renormalized
local temperature
\begin{equation}
\delta \tilde T\equiv\delta T-\frac{2}{3}T_0\frac{\sigma_1\tau_{\mu}}{\tilde n_0}
\delta\Gamma_{12},
\label{reT}
\end{equation}
and a renormalized local chemical potential
\begin{equation}
\delta\tilde\mu\equiv\delta \mu_c^{(0)}-g(\sigma_H^{-1}+1)\tau_{\mu}\delta\Gamma_{12}.
\label{remu}
\end{equation}
Using $\delta\mu_c=\delta\mu_c^{(0)}-g\tau_{\mu}\delta\Gamma_{12}$
(see (\ref{muctot})) and $\delta \tilde \mu$ from (\ref{remu}), we obtain
\begin{equation}
\delta\mu_{\rm diff}\equiv\delta\tilde\mu-\delta\mu_c
=\delta\tilde \mu-[\delta\mu_c^{(0)}-g\tau_{\mu}
\delta\Gamma_{12}]=g\sigma_H^{-1}\delta\Gamma_{12}.
\end{equation}
This relation between $\delta\mu_{\rm diff}$ and $\delta\Gamma_{12}$ is precisely
that given by (\ref{G12}) and (\ref{taumu}), as derived in the ZGN$'$ theory.
The physical significance of the renormalized thermodynamic quantities, as
given by (\ref{reT}) and (\ref{remu}), will become clear in Section \ref{sec:LK}. 

\section{Equivalence to Landau-Khalatnikov Two-Fluid Equations with dissipation}
\label{sec:LK}

In this Section, we prove that our hydrodynamic equations in (\ref{hydro-C})
and (\ref{hydro-n}) can be written in the form of the Landau-Khalatnikov two-fluid equations.
We first display the complete Landau-Khalatnikov two-fluid equations involving
dissipative terms~\cite{Khal}:
\begin{mathletters}
\begin{equation}
{\partial n\over\partial t} + \bbox{\nabla}\cdot {\bf j}=0, 
\label{eq_n}
\end{equation}
\begin{eqnarray}
&&m\frac{\partial j_{\mu}}{\partial t}+\frac{\partial}{\partial x_{\nu}}
(\delta_{\mu\nu}P+m\tilde n v_{n\mu}v_{n\nu}+mn_c v_{c\mu}v_{c\nu})
+n\frac{\partial U_{ext}}{\partial x_{\mu}} \cr
&& ~~~~~=\frac{\partial}{\partial x_{\nu}}\left\{
 2\eta\left[D_{\mu\nu}-\frac{1}{3}\delta_{\mu\nu}({\rm Tr}D)\right]
 +\delta_{\mu\nu}(\zeta_1\bbox{\nabla}\cdot[mn_c({\bf v}_c-{\bf v}_n)]
 +\zeta_2\bbox{\nabla}\cdot{\bf v}_n)\right\},
\label{eq_j}
\end{eqnarray}
\begin{equation}
{\partial{\bf v}_c\over\partial t} = -\bbox{\nabla}
\left\{\frac{\mu}{m}+\frac{v_c^2}{2}-\zeta_3\bbox{\nabla}\cdot
[mn_c({\bf v}_c-{\bf v}_n)]-\zeta_4\bbox{\nabla}\cdot{\bf v}_n\right\},
\label{eq_vs}
\end{equation}
\begin{equation}
{\partial s\over\partial t} +\bbox{\nabla}\cdot\left(s{\bf v}_n-
\frac{\kappa\bbox{\nabla}T}{T}\right)=\frac{R_s}{T}.
\label{eq_s}
\end{equation}
\label{landau_eqs}
\end{mathletters}

\noindent
The total current is given by
${\bf j}= n_c{\bf v}_c+ \tilde n {\bf v}_n$ and the dissipative function describing the
entropy production rate is given by \cite{Khal}
\begin{eqnarray}
R_s&=&\zeta_2(\bbox{\nabla}\cdot{\bf v}_n)^2+2\zeta_1\bbox{\nabla}\cdot{\bf v}_n
\bbox{\nabla}\cdot[mn_c({\bf v}_c-{\bf v}_n)]+
\zeta_3\left(\bbox{\nabla}\cdot[mn_c({\bf v}_c-{\bf v}_n)]\right)^2 \cr
&&+2\eta\left[D_{\mu\nu}-\frac{1}{3}\delta_{\mu\nu}({\rm Tr}D)\right]^2
+\frac{\kappa}{T}(\bbox{\nabla} T)^2 .
\label{s-rate}
\end{eqnarray}
As we have discussed in Ref~\cite{ZNG}, the normal fluid and the superfluid
densities that appear in the standard Landau two-fluid theory can be identified with
the corresponding non-condensate and condensate densities, within the context
of our finite temperature model based on the HF approximation for single-particle excitations.
We have explicitly made use of this correspondence in writing (\ref{landau_eqs}).
We also note that in (\ref{landau_eqs}) and (\ref{s-rate}),
one can write $n_c({\bf v}_c-{\bf v}_n)$ in the
equivalent form $({\bf j}-n{\bf v}_n)$, which is often used.		

The thermodynamic functions that appear in these Landau-Khalatnikov two-fluid equations
satisfy the following superfluid local thermodynamic relations:
\begin{mathletters}
\begin{eqnarray}
&&P+\epsilon=\mu n+sT+m\tilde n({\bf v}_n-{\bf v}_c)^2, 
\label{relation_a} \\
&&dP=nd\mu +sdT-m\tilde n({\bf v}_n-{\bf v}_c)\cdot d({\bf v}_n-{\bf v}_c), 
\label{relation_b} \\
&&d\epsilon=\mu dn +Tds+({\bf v}_n-{\bf v}_c)\cdot d[m\tilde n({\bf v}_n-{\bf v}_c)].
\label{relation_c}
\end{eqnarray}
\label{thermo-relation}
\end{mathletters}

\noindent
The various local thermodynamic functions which appear in the LK theory have to be
carefully defined so that they satisfy the relations in (\ref{thermo-relation}).
The local entropy is defined by (as in Ref.~\cite{ZNG})
\begin{equation}
s=\int\frac{d{\bf p}}{(2\pi\hbar)^3}
\left[(1+f)\ln(1+f)-f\ln f\right].
\end{equation}
Using $f=f^{(0)}+f^{(1)}$ and working to first order in $f^{(1)}$, one finds
\begin{equation}
s=\int\frac{d{\bf p}}{(2\pi\hbar)^3}
\left[(1+f^{(0)})\ln(1+f^{(0)})-f^{(0)}\ln f^{(0)}+
\ln(1+f^{(0)^{-1}})f^{(1)}\right].
\end{equation}
From the constraint on $f^{(1)}$ given by (\ref{const_s}), one sees that
the last term, which arises from $f^{(1)}$, makes no contribution to the
local entropy. 
One thus obtains
\begin{equation}
s=
\frac{1}{T}\left[\frac{5}{2}\tilde P^{(0)}-\tilde n^{(0)}
(\tilde \mu^{(0)}-U)\right]
\simeq
\frac{1}{T}\left[\frac{5}{2}\tilde P^{(0)}-\tilde n^{(0)}
(\mu_c^{(0)}-U)-\frac{m\tilde n}{2}({\bf v}_n-{\bf v}_c)^2\right],
\end{equation}
where we have used $\tilde n^{(0)}=\tilde n+O(v_n,v_c)$.

The local energy density $\epsilon$ in the Landau-Khalatnikov theory is defined in the
local frame where ${\bf v}_c=0$~\cite{HM}.
In the context of the present theory, this is given by
\begin{equation}
\epsilon=\tilde \epsilon+nU_{\rm ext}+\frac{g}{2}(n^2+2n\tilde n-\tilde n^2)
+\frac{m}{2}\tilde n({\bf v}_n-{\bf v}_c)^2,
\label{totalenergy}
\end{equation}
while the local energy density in the original lab frame is given by
\begin{equation}
\epsilon_{\rm lab}=\epsilon+m\tilde n({\bf v}_n-{\bf v}_c)\cdot{\bf v}_c
+\frac{mn}{2}v_c^2.
\end{equation}
Using (\ref{ntildetot}) and (\ref{ptildetot}) in (\ref{totalenergy}),
one finds that the first order corrections from $\Gamma_{12}$ cancel out, leaving
\begin{equation}
\epsilon=\tilde \epsilon^{(0)}+nU_{\rm ext}+\frac{g}{2}(n^2+2n\tilde n^{(0)}-[\tilde n^{(0)}]^2)
+\frac{1}{2}m\tilde n({\bf v}_n-{\bf v}_c)^2.
\end{equation}
We conclude that {\it both} the local entropy density and the local energy density are
determined by the diffusive local equilibrium distribution function $f^{(0)}$ alone,
and are not altered by the deviation $f^{(1)}$.

In contrast, as we now show, the total pressure and the chemical potential must be carefully 
defined so that they satisfy the superfluid thermodynamic relations in (\ref{thermo-relation}).
We first define the {\it non-equilibrium} pressure by
\begin{equation}
P'\equiv\tilde P+\frac{g}{2}(n^2+2n\tilde n-\tilde n^2).
\label{neqP}
\end{equation}
Using (\ref{ntildetot}) and (\ref{ptildetot}) and working to first
order in $\Gamma_{12}^{(1)}$, one obtains
\begin{equation}
P'=P-\tau_{\mu}\frac{gn_c}{3}\Gamma_{12}^{(1)},
\label{neqP2}
\end{equation}
where $P$ is the (diffusive) local equilibrium pressure defined as
\begin{equation}
P\equiv\tilde P^{(0)}+\frac{g}{2}(n^2+2n\tilde n^{(0)}-[\tilde n^{(0)}]^2). \\
\label{eqP}
\end{equation}
We find that the LK thermodynamic relations given in (\ref{thermo-relation}) are not
satisfies if we assume that $P'$ is the pressure ($P$) and $\mu_c$ is the
chemical potential ($\mu$).
Extra terms appear which are associated with $\Gamma_{12}$.
This means that the above identification of the thermodynamic variables is only valid in the
lowest order hydrodynamic equations, where there is no dissipation.
 
We recall that in deriving the Landau equations from the ZGN$'$ equations
in Ref.~\cite{ZNG}, we defined the total pressure by (\ref{neqP}) and $\mu=\mu_c$,
and also found extra terms in the thermodynamic relations proportional to
$\delta\mu_{\rm diff}$ (see Eq.~(71) of Ref.~\cite{ZNG}).
Therefore the precise equivalence between the ZGN$'$ hydrodynamics and the Landau
theory shown in Ref.~\cite{ZNG} was restricted in the limit $\omega\tau_{\mu}\to 0$,
i.e., when $\delta\mu_{\rm diff}\to 0$.
In contrast, if the pressure $P$ is defined to be (\ref{eqP}) and $\mu=\mu_c^{(0)}$, 
we can show that the the superfluid thermodynamic relations in (\ref{thermo-relation})
are satisfied.
Therefore we conclude that the local equilibrium pressure defined in (\ref{eqP})
and the local equilibrium chemical potential $\mu_c^{(0)}$ given by (\ref{muc01})
are, in fact, the correct variables to be use in the Landau-Khalatnikov equations.
We will show later that the corrections to the total pressure and the
chemical potential actually give rise to the additional damping terms associated with
the four second viscosity coefficients $\zeta_i$ in (\ref{landau_eqs}).

We now proceed to derive the LK equations from our microscopic theory, one by one.
Our continuity equations for $n_c$ and $\tilde n$ are given by
(\ref{eq_nc}) and (\ref{eq_tilden}).
Adding them, we obtain the continuity equation for the total
density (\ref{eq_n}).
To derive the equation (\ref{eq_j}) for the total current ${\bf j}$,
we combine our two continuity equations and the two velocity
equations (\ref{eq_vc}) and (\ref{eq_vn}) to give
\begin{eqnarray}
&&m\frac{\partial j_{\mu}}{\partial t}+\frac{\partial}{\partial x_{\nu}}
(\delta_{\mu\nu}P'+m\tilde n v_{n\mu}v_{n\nu}+mn_c v_{c\mu}v_{c\nu})
+n\frac{\partial U_{\rm ext}}{\partial x_{\mu}} \cr
&& =\frac{\partial}{\partial x_{\nu}}\left\{
 2\eta\left[D_{\mu\nu}-\frac{1}{3}\delta_{\mu\nu}({\rm Tr}D)\right]\right\}.
\label{eq_j2}
\end{eqnarray}
Using (\ref{neqP2}), we find
\begin{eqnarray}
&&m\frac{\partial j_{\mu}}{\partial t}+\frac{\partial}{\partial x_{\nu}}
(\delta_{\mu\nu}P+m\tilde n v_{n\mu}v_{n\nu}+mn_c v_{c\mu}v_{c\nu})
+n\frac{\partial U_{\rm ext}}{\partial x_{\mu}} \cr
&& =\frac{\partial}{\partial x_{\nu}}\left\{
 2\eta\left[D_{\mu\nu}-\frac{1}{3}\delta_{\mu\nu}({\rm Tr}D)\right]
+\delta_{\mu\nu}\tau\frac{gn_c}{3}\Gamma_{12}^{(1)}\right\}.
\label{eq_j3}
\end{eqnarray}
To consistently include damping due to the first order correction
term in the Chapman-Enskog expansion, we use
$\Gamma_{12}=\Gamma_{12}^{(1)}$ as given in (\ref{gamma1}).
We then find that (\ref{eq_j3}) is identical with the LK equation (\ref{eq_j})
with the second viscosity coefficients $\zeta_1$ and $\zeta_2$ 
given by
\begin{equation}
\zeta_1=\frac{gn_c}{3m}\tau_{\mu}\sigma_H, ~~ 
\zeta_2=\frac{gn_c^2}{9}\tau_{\mu}\sigma_H.
\label{zeta12}
\end{equation}
Using $\mu_c=\mu_c^{(0)}+g\tilde n^{(1)}=\mu-g\tau\Gamma_{12}^{(1)}$ (see (\ref{muc01}))
and the expression for $\Gamma_{12}^{(1)}$ in our equation for the
condensate velocity given in (\ref{eq_vc}), we find the latter can be written precisely
in the LK form (\ref{eq_vs}).
Comparison between the two equations shows that the second viscosity coefficients
$\zeta_3$ and $\zeta_4$ are
given by
\begin{equation}
\zeta_3=\frac{g}{m^2}\tau_{\mu}\sigma_H, ~~\zeta_4=\frac{gn_c}{3m}\tau_{\mu}\sigma_H.
\label{zeta34}
\end{equation}
We note that our results for the second viscosities satisfy the Onsager reciprocal
relation $\zeta_1=\zeta_4$ (this equality follows quite generally, as shown
by Eq.~(4.28) of Ref.~\cite{HM}).

Finally, we derive the equation for the local entropy.
Using (\ref{relation_c}), we have
\begin{equation}
T\frac{\partial s}{\partial t}=\frac{\partial \epsilon}{\partial t}
-\mu\frac{\partial n}{\partial t}
-m({\bf v}_n-{\bf v}_c)\frac{\partial}{\partial t}
[\tilde n({\bf v}_n-{\bf v}_c)].
\label{entropy_1}
\end{equation}
With the expression for the local energy density $\epsilon$ given in (\ref{totalenergy}),
we find (\ref{entropy_1}) reduces to
\begin{eqnarray}
\frac{\partial s}{\partial t}&=&\frac{\partial\tilde\epsilon}{\partial t}
+[U_{\rm ext}-\mu+g(n+\tilde n)]\frac{\partial n}{\partial t}
+gn_c\frac{\partial \tilde n}{\partial t}-\frac{\partial n}{\partial t}
\frac{m}{2}({\bf v}_n-{\bf v}_c)^2 \cr
&& \cr
&=&\frac{\partial\tilde\epsilon}{\partial t}+g\tilde n^{(1)}\frac{\partial n}{\partial t}
+gn_c\frac{\partial \tilde n}{\partial t}.
\label{entropy_2}
\end{eqnarray}
Here we have neglected the last term in the first line, since it is
of third order in the local velocities.
Using our hydrodynamic equations (\ref{eqhydro}), 
we find (\ref{entropy_2}) reduces to the form (\ref{eq_s}),
assuming the entropy production rate $R_s$ is given by
\begin{eqnarray}
R_s&=&\tau_{\mu} g\Gamma_{12}
\left\{g\bbox{\nabla}\cdot[n_c({\bf v}_c-{\bf v}_n)]+\frac{1}{3}n_c\bbox{\nabla}\cdot{\bf v}_n
\right\} \cr
&&+2\eta\left[D_{\mu\nu}-\frac{1}{3}\delta_{\mu\nu}({\rm Tr}D)\right]^2
+\frac{\kappa}{T}(\bbox{\nabla} T)^2 .
\label{s-rate2}
\end{eqnarray}
Using (\ref{gamma1}) and the expression in (\ref{zeta12}) and (\ref{zeta34}),
we see that (\ref{s-rate2}) is equivalent to the Landau-Khalatnikov~\cite{Khal}
expression given in (\ref{s-rate}).

We have thus shown that our equations based on a microscopic theory built on Bose
condensation can be written in a form precisely identical
to the phenomenological Landau-Khalatnikov two-fluid equations including
the damping associated with the shear viscosity, 
thermal conductivity and four second viscosity coefficients.
An analogous derivation of the Landau-Khalatnikov equations
for a uniform Bose gas was first given by Kirkpatrick and Dorfman \cite{KD}.
However at finite temperatures, where the dominant excitations are particle-like
Hartree-Fock excitations, KD did not obtain the second viscosities since they neglected
the source term
$\Gamma_{12}$ associated with deviation from local equilibrium produced by the 
$C_{12}$ collisions.
We have shown that the second viscosity coefficients are directly related to the 
$\Gamma_{12}$ term first discussed in Ref.~\cite{ZNG},
which represents the collisional exchange of atoms between
the condensate and non-condensate.

In the above derivation of the second viscosity terms,
we used $\Gamma_{12}=\Gamma_{12}^{(1)}$.
This restricts the validity of the results to the case $\omega\tau_{\mu}\ll 1$
when we consider collective fluctuations with frequency $\omega$.
However, our discussion can be easily extended to the situation when $\omega\tau_{\mu}$ is
not small, by using (see (\ref{eq_mudiff2}))
\begin{equation}
\Gamma_{12}(\omega)=\frac{\sigma_H}{1-i\omega\tau_{\mu}}
\left\{\bbox{\nabla}\cdot[n_c({\bf v}_c-{\bf v}_n)]+\frac{1}{3}n_c\bbox{\nabla}\cdot
{\bf v}_n\right\}.
\end{equation}
Using this expression, we can still write our equations in the Landau-Khalatnikov form,
but now with the {\it frequency-dependent} second viscosity coefficients
\begin{equation}
\zeta_i(\omega)=\frac{\zeta_i}{1-i\omega\tau_{\mu}}.
\label{zeta_omega}
\end{equation}
Everything else in our derivation goes through.

The expression (\ref{zeta_omega}) for the frequency-dependent second viscosity coefficients has
in fact the expected form, as derived from general considerations \cite{LL}.
The second viscosity , such as associated with compression and expansion, 
arises when a gas is coupled
to an internal relaxation process (for example, the transfer of energy from
the translational degrees of freedom of a molecule to the vibrational degrees of
freedom).
If the relaxation time of the internal process is denoted by $\tau_{\rm R}$,
the frequency-dependent second viscosity coefficient is given by
$\zeta(\omega)=\zeta_0/(1-i\omega\tau_{\rm R})$, 
where $\zeta_0\propto \tau_{\rm R}$.
In a Bose-condensed gas at finite temperatures, the non-condensate atoms are
coupled to the condensate degree of freedom, and we have shown that 
the characteristic relaxation
time for the equilibration between the two components is given by $\tau_{\mu}$.
In this connection, we might recall that
in the superfluid $^4$He, the second viscosities are due to the fact that
the total number of phonons and rotons is not conserved~\cite{Khal}.
Above $T_{\rm BEC}$ (when $n_c=0$), all the second viscosity coefficients $\zeta_i$
in (\ref{zeta12}) and (\ref{zeta34}) vanish, as expected in a normal dilute
single-component gas.

We finally note that the Landau-Khalatnikov equations could have been
derived from the ZGN$'$ hydrodynamic equations if we simply identified the
total pressure $P$ with that by (\ref{neqP}) and (\ref{neqP2}),
and took the chemical potential to be $\mu=\mu_c+g\tau_{\mu}\delta\Gamma_{12}$
(see Eq.~(\ref{muctot})).
This leads more naturally to frequency-dependent second viscosities.
On the other hand, the physical meaning of this choice of
the local equilibrium pressure and  chemical potential is not made clear.

\section{Calculation of transport coefficients}
\label{sec:transport}
In this section, we solve the linearized equation for functions $A$, $B$
and $D$ in (\ref{eqforABD}) which determine the deviation from local equilibrium
as described by $\psi$ in (\ref{correction}) and (\ref{solution}).
We can then calculate the transport coefficients $\eta$ and $\kappa$ as given
in (\ref{eta-kappa}).
We follow the standard procedure in the Chapman-Enskog method, as reviewed in \cite{ferziger}.
In this approach, one solves the linearized equation
by expanding $\psi$ in a basis set of polynomial functions.
These polynomial functions are chosen to satisfy the constraints such as (\ref{const}) which
the solution $\psi$ must satisfy.
In a classical gas, one uses Sonine polynomials \cite{ferziger}.
One can also define analogous polynomials for a degenerate Bose gas~\cite{UU}.
As usual, we calculate the transport coefficients using
the lowest order polynomial approximation, which 
usually gives very accurate results for
the transport coefficients.
For a more detailed mathematical discussion which is easily generalized to
Bose condensed gases, we refer to \cite{ferziger,UU}.

\subsection{The thermal conductivity}
In evaluating the thermal conductivity, it is convenient to rewrite
(\ref{kappa}) as
\begin{eqnarray}
\kappa&=&-\frac{1}{3}k_{\rm B}\int\frac{d{\bf p}}{(2\pi\hbar)^3}{\bf u}A(u)
\cdot{\bf u}\left[\frac{mu^2}{2k_{\rm B}T_0}-\frac{5g_{5/2}(z_0)}{2g_{3/2}(z_0)}\right]
f_0(1+f_0) \cr
&=&-\frac{1}{3}k_{\rm B}\int\frac{d{\bf p}}{(2\pi\hbar)^3}{\bf u}A(u)
\cdot\hat L[{\bf u}A(u)].
\label{kappa2}
\end{eqnarray}
Here we explicitly use the static equilibrium distribution function $f_0$
to evaluate the transport coefficients.
To solve the linear integral equation (\ref{eqforA}) for $A(u)$, we introduce a
simple ansatz of the form \cite{NG,UU}:
\begin{equation}
A(u)=A\left[\frac{mu^2}{2k_{\rm B}T_0}-\frac{5g_{5/2}(z_0)}{2g_{3/2}(z_0)}\right].
\label{ansatzA}
\end{equation} 
This is the lowest order polynomial function that satisfies the constraint
given by (\ref{constA}).
The constant $A$ in (\ref{ansatzA}) is determined by multiplying (\ref{eqforA}) by 
${\bf u}[mu^2/2k_{\rm B}T-5g_{5/2}(z_0)/2g_{3/2}(z_0)]$ and integrating
over ${\bf p}$, giving
\begin{eqnarray}
A&=&
\int \frac{d{\bf p}}{(2\pi\hbar)^3} u^2
\left[\frac{mu^2}{2k_{\rm B}T_0}-\frac{5g_{5/2}(z_0)}{2g_{3/2}(z_0)}\right]^2
f_0(1+f_0) \cr
& &
\times
\left\{\int \frac{d{\bf p}}{(2\pi\hbar)^3}
\left[\frac{mu^2}{2k_{\rm B}T_0}-\frac{5g_{5/2}(z_0)}{2g_{3/2}(z_0)}
\right]{\bf u}
\cdot \hat L\left[
\left\{\frac{mu^2}{2k_{\rm B}T_0}-\frac{5g_{5/2}(z_0)}{2g_{3/2}(z_0)}
\right\}{\bf u}\right]\right\}^{-1}
\label{eqAconst}
\end{eqnarray}
Using (\ref{ansatzA}) and (\ref{eqAconst}) in (\ref{kappa2}), we find
\begin{eqnarray}
\kappa&=&-\frac{k_{\rm B}}{3}
\left\{\int \frac{d{\bf p}}{(2\pi\hbar)^3} u^2
\left[\frac{mu^2}{2k_{\rm B}T_0}-\frac{5g_{5/2}(z_0)}{2g_{3/2}(z_0)}\right]^2
f_0(1+f_0) \right\}^2\cr
& &
\times
\left\{\int \frac{d{\bf p}}{(2\pi\hbar)^3}
\left[\frac{mu^2}{2k_{\rm B}T_0}-\frac{5g_{5/2}(z_0)}{2g_{3/2}(z_0)}
\right]{\bf u}
\cdot \hat L\left[
\left\{\frac{mu^2}{2k_{\rm B}T_0}-\frac{5g_{5/2}(z_0)}{2g_{3/2}(z_0)}
\right\}{\bf u}\right]\right\}^{-1}. 
\label{kappa3}
\end{eqnarray}

To evaluate the ${\bf p}$ integrals in (\ref{kappa3}), it is convenient to introduce
the dimensionless velocity variable by
\begin{equation}
{\bf x}\equiv\left(\frac{m}{2k_{\rm B}T_0}\right)^{1/2}{\bf u}=
\left(\frac{1}{2mk_{\rm B}T_0}\right)^{1/2}{\bf p}.
\label{u2x}
\end{equation}
With this new variable, we can rewrite the linearized collision operator as
\begin{equation}
\hat L[\psi]=\frac{8m(k_{\rm B}T_0)^2a^2}{\pi^3\hbar^3}
\left(\hat L'_{22}[\psi]+\pi^{3/2}n_{c0}\Lambda^3_0\hat L_{12}'[\psi]\right),
\end{equation}
where the dimensionless collision operators $\hat L_{22}'$ and $\hat L_{12}'$ 
are defined by
\begin{eqnarray}
\hat L_{22}'[\psi]&\equiv&\int d{\bf x}_2\int d{\bf x}_3\int d{\bf x}_4
\delta({\bf x}+{\bf x}_2-{\bf x}_3-{\bf x}_4)
\delta(x^2+x_2^2-x_3^2-x_4^2) \cr
&&\times f_{10}f_{20}(1+f_{30})(1+f_{40})
(\psi_3+\psi_4-\psi_2-\psi),
\end{eqnarray}
\begin{eqnarray}
\hat L_{12}'[\psi]&\equiv&\int d{\bf x}_1\int d{\bf x}_2\int d{\bf x}_3
\delta({\bf x}-{\bf x}_2-{\bf x}_3)
\delta(x_1^2-\beta_0 gn_{c0}-x_2^2-x_3^2) \cr
&&\times [\delta({\bf x}-{\bf x}_1)-\delta({\bf x}-{\bf x}_2)-\delta({\bf x}-{\bf x}_3)]
(1+f_{10})f_{20}f_{30}(\psi_2+\psi_3-\psi),
\end{eqnarray}
where $f_{i0}=(z_0^{-1}e^{x_i^2}-1)^{-1}$.
Carrying out the ${\bf p}$ (or ${\bf u}$) integrals in (\ref{kappa3}), one finds
\begin{equation}
\kappa=\frac{75k_{\rm B}}{64a^2m}\left(\frac{mk_{\rm B}T}{\pi}\right)^{1/2}
\frac{\pi^{1/2}}{I^{\kappa}_{22}(z_0)+\Lambda^3_0n_{c0}I_{12}^{\kappa}(z_0)}
\left[\frac{7}{2}g_{7/2}(z_0)-\frac{5g^2_{5/2}(z_0)}{2g_{3/2}(z_0)} \right]^2,
\label{kappa_final}
\end{equation}
where the functions $I_{22}^{\kappa}(z_0)$ and $I_{12}^{\kappa}(z_0)$ are defined by
\begin{equation}
I_{22}^{\kappa}\equiv-\int d{\bf x}~{\bf x}x^2\cdot\hat L_{22}'[{\bf x}x^2],
\end{equation}
\begin{equation}
I_{12}^{\kappa}\equiv-\pi^{3/2}\int d{\bf x}~{\bf x}x^2\cdot\hat L_{22}'[{\bf x}x^2].
\end{equation}

In Ref.~\cite{NG}, we derived a convenient
formula for the integral $I_{22}^{\kappa}$, namely
\begin{eqnarray}
I_{22}^{\kappa}(z_0)&=&\sqrt{2}\pi^3=\int_0^{\infty}dx_0\int_0^{\infty}dx_r
\int_{-1}^1dy\int_{-1}^1 dy' F_{22}(x_0,x_r,y,y';z_0) \cr
&&\times x_0^4x_r^7(y^2+y'^2-2y^2y'^2),
\end{eqnarray}
where
\begin{equation}
F_{22}(x_0,x_r,y,y';z_0)
=\frac{z^2_0e^{-x_0^2-x_r^2}}
{(1-z_0e^{-x_1^2})(1-z_0e^{-x_2^2})(1-z_0e^{-x_3^2})(1-z_0e^{-x_4^2})},
\label{F22}
\end{equation}
with
\begin{eqnarray}
&&x_1^2=\frac{1}{2}(x_0^2+2x_0x_ry+x_r^2), ~x_2^2=\frac{1}{2}(x_0^2-2x_0x_ry+x_r^2) \cr
&&x_3^2=\frac{1}{2}(x_0^2+2x_0x_ry'+x_r^2), ~x_4^2=\frac{1}{2}(x_0^2-2x_0x_ry'+x_r^2).
\end{eqnarray}
We note that $I_{22}^{\kappa}(z_0)$ is a universal function of the equilibrium fugacity
$z_0$, where $z_0=e^{-\beta gn_{c0}({\bf r})}$.
To derive a similar expression for $I_{12}^{\kappa}$, we introduce the
transformation 
\begin{equation}
{\bf x}_2=\frac{1}{2}({\bf x}_0+{\bf x}_r), ~~
{\bf x}_3=\frac{1}{2}({\bf x}_0-{\bf x}_r).
\end{equation}
We then express ${\bf x}_r$ in the polar coordinate $(x_r,\theta,\phi)$
where $\theta$ is the azimuthal angle with respect to the vector ${\bf x}_0$,
ie, ${\bf x}_r\cdot{\bf x}_0=x_rx_0\cos\theta$.
With these new variables, one obtains the following expression for $I_{12}^{\kappa}$:
\begin{eqnarray}
I_{12}^{\kappa}(z_0)&=&8\pi^{7/2}\int_0^{\infty}dx_r\int_{-1}^1 dy F_{12}(x_r,y;z_0)
x_r^2(x_r^2+\beta_0 gn_{c0})^{3/2} \cr
&&\times\left[x_r^2(x_r^2+3\beta_0 gn_{c0})(1-y^2)+\frac{9}{4}(\beta_0 gn_{c0})^2\right],
\end{eqnarray}
where $y=\cos\theta$ and
\begin{equation}
F_{12}(x_r,y;z_0)=\frac{ z_0e^{-x_1^2} }
{ (1-z_0e^{-x_1^2})(1-z_0e^{-x_2^2})(1-z_0e^{-x_3^2}) },
\label{F12}
\end{equation}
with
\begin{eqnarray}
&&x_1^2=2(x_r^2+\beta_0 gn_{c0}), \cr
&&x_2^2=x_r^2+x_ry\sqrt{x_r^2+\beta_0 gn_{c0}}+\frac{1}{2}\beta_0 gn_{c0} \cr
&&x_3^2=x_r^2-x_ry\sqrt{x_r^2+\beta_0 gn_{c0}}+\frac{1}{2}\beta_0 gn_{c0}.
\end{eqnarray}
The formula in (\ref{kappa_final}) gives the thermal conductivity $\kappa$
as a universal function of $gn_{c0}({\bf r})$ or equivalently in terms of
the local fugacity $z_0({\bf r})=e^{-\beta_0 gn_{c0}({\bf r})}$.
If we ignore the contribution from $C_{12}$ collisions, i.e.,
set $I_{12}^{\kappa}$ to zero, (\ref{kappa_final}) reduces to the
expression for $\kappa$ derived in our earlier work~\cite{NG,CJP}.

One can also write the expression for $\kappa$ in (\ref{kappa_final}) in the
following useful form:
\begin{equation}
\kappa=\frac{5\sqrt{2}}{\pi^3}
\left(\tau_{\kappa}\frac{\tilde n_0 k_{\rm B}^2T_0}{m}\right)
\left\{\frac{7g_{7/2}(z_0)}{2g_{5/2}(z_0)}-
\left[\frac{5g_{5/2}(z_0)}{2g_{3/2}(z_0)}\right]^2 \right\},
\label{kappa_final2}
\end{equation}
where $\tau_{\kappa}$ is the ``thermal relaxation time''
associated with the thermal conductivity, as defined in Appendix \ref{relax_times}.
In turn, one can also write the reciprocal of this relaxation time $\tau_{\kappa}$ as
the sum of contributions from $C_{12}$ and $C_{22}$ collisions,
\begin{equation}
\frac{1}{\tau_{\kappa}}=\frac{1}{\tau_{\kappa,12}}+\frac{1}{\tau_{\kappa,22}},
\label{tau_k2}
\end{equation}
where these relaxation times are given explicitly in Appendix~\ref{relax_times}.
The physical meaning of this $\tau_{\kappa}$ relaxation time is discussed in
Appendix \ref{relax_times}, using a simple relaxation time approximation
for the collision integrals in (\ref{eq1}).

\subsection{The shear viscosity}
In evaluating the shear viscosity $\eta$, it is convenient to rewrite (\ref{eta}) as
\begin{eqnarray}
\eta&=&
-\frac{m}{10}\int \frac{d{\bf p}}{(2\pi\hbar)^3}
\left(u_{\mu}u_{\nu}-\frac{1}{3}\delta_{\mu\nu}u^2\right)
B(u)\left(u_{\mu}u_{\nu}-\frac{1}{3}\delta_{\mu\nu}u^2\right)
f_0(1+f_0) \cr
&=&
-\frac{k_{\rm B}T_0}{10}\int \frac{d{\bf p}}{(2\pi\hbar)^3}
B(u)\left(u_{\mu}u_{\nu}-\frac{1}{3}\delta_{\mu\nu}u^2\right)
\hat L\left[B(u)\left(u_{\mu}u_{\nu}-\frac{1}{3}\delta_{\mu\nu}u^2\right)\right].
\label{eta1}
\end{eqnarray}
To solve (\ref{eta1}), the simplest consistent ansatz \cite{NG,UU} is to use $B(u)\equiv B$.
The constant $B$ can be determined by multiplying (\ref{eqforB}) by
$(u_{\mu}u_{\nu}-\delta_{\mu\nu}u^2/3)$ and integrating over ${\bf p}$, to give
\begin{eqnarray}
B&=&\frac{m}{k_{\rm B}T_0}
\left\{\int \frac{d{\bf p}}{(2\pi\hbar)^3}
\left(u_{\mu}u_{\nu}
-\frac{1}{3}\delta_{\mu\nu}u^2\right)^2
f_0(1+f_0)\right\} \cr
&&\times\left\{\int \frac{d{\bf p}}{(2\pi\hbar)^3}
\left(u_{\mu}u_{\nu}-\frac{1}{3}\delta_{\mu\nu}u^2\right)
\hat L\left[u_{\mu}u_{\nu}
-\frac{1}{3}\delta_{\mu\nu}u^2 \right]\right\}^{-1}.
\label{eqBconst}
\end{eqnarray}
Using this in (\ref{eta1}), we obtain
\begin{eqnarray}
\eta&=&-\frac{m^2}{10k_{\rm B}T_0}
\left\{\int \frac{d{\bf p}}{(2\pi\hbar)^3}
\left(u_{\mu}u_{\nu}
-\frac{1}{3}\delta_{\mu\nu}u^2\right)^2
f_0(1+f_0)\right\}^2 \cr
&&\left\{\int \frac{d{\bf p}}{(2\pi\hbar)^3}
\left(u_{\mu}u_{\nu}-\frac{1}{3}\delta_{\mu\nu}u^2\right)
\hat L\left[u_{\mu}u_{\nu}
-\frac{1}{3}\delta_{\mu\nu}u^2 \right]\right\}^{-1}.
\label{eta2}
\end{eqnarray}
With the dimensionless variable defined in (\ref{u2x}), this expression for
the shear viscosity $\eta$ can be rewritten as
\begin{equation}
\eta=\frac{5\pi^3}{32\sqrt{2}a^2}
(mk_{\rm B}T_0)^{1/2}
\frac{g^2_{5/2}(z_0) }{I_{22}^{\eta}(z_0)+\Lambda^3_0 n_{c0} I_{12}^\eta(z_0) },
\label{etafinal}
\end{equation}
where
\begin{eqnarray}
I_{22}^{\eta}&\equiv& -\int d{\bf x}
\left(x_{\mu}x_{\nu}-\frac{1}{3}\delta_{\mu\nu}x^2\right)
\hat L'_{22}\left[x_{\mu}x_{\nu}
-\frac{1}{3}\delta_{\mu\nu}x^2 \right] \cr
&=&\frac{\pi^3}{\sqrt{2}}\int_0^{\infty} dx_0 \int_0^{\infty}
dx_r\int_{-1}^1 dy \int_{-1}^1 dy' F_{12}(x_0,x_r,y,y';z_0) \cr
&& \times x_0^2x_r^7(1+y^2+y'^2-3y^2y'^2),
\end{eqnarray}
and
\begin{eqnarray}
I_{12}^{\eta}&\equiv& -\pi^{3/2}\int d{\bf x}
\left(x_{\mu}x_{\nu}-\frac{1}{3}\delta_{\mu\nu}x^2\right)
\hat L'_{12}\left[x_{\mu}x_{\nu}
-\frac{1}{3}\delta_{\mu\nu}x^2 \right] \cr
&=&8\pi^{7/2}\int_0^{\infty}dx_r\int_{-1}^{1}dy
F_{12}(x_r,y;z_0) \cr
&&\times x_r^2\sqrt{x_r^2+\beta_0 gn_{c0}}
\left[x_r^2(x_r^2+\beta_0 gn_{c0})(1-y^2)+\frac{1}{3}(\beta_0 gn_{c0})^2\right].
\end{eqnarray}
These expressions involve the same functions $F_{22}$ and $F_{12}$ defined 
earlier in (\ref{F22}) and (\ref{F12}).

Analogous to (\ref{kappa_final2}) for the thermal conductivity,
can also write the expression for $\eta$ in (\ref{etafinal}) in the
following form:
\begin{equation}
\eta=\tau_{\eta}\tilde n_0k_{\rm B}T_0 \left[\frac{g_{5/2}(z_0)}{g_{3/2}(z_0)}\right],
\label{etafinal2}
\end{equation}
where the viscous relaxation time $\tau_{\eta}$ is defined in Appendix~\ref{relax_times}.
Again one can write the reciprocal of the relaxation time $\tau_{\eta}$ as
\begin{equation}
\frac{1}{\tau_{\eta}}=\frac{1}{\tau_{\eta,12}}+\frac{1}{\tau_{\eta,22}},
\label{tau_e2}
\end{equation}
where these $C_{12}$ and $C_{22}$
relaxation times are given in Appendix~\ref{relax_times}.
As with $\tau_{\kappa}$, the relaxation time $\tau_{\eta}$ can be understood in
terms of a simple relaxation time approximation for the collision term in the
kinetic equation.

\subsection{The second viscosity coefficients}
To find the expression for
$\tau$ as defined in (\ref{tau}), we use the simple
ansatz for the form of the solution for $D(u)$ of (\ref{eqforD}),
\begin{equation}
D(u)=D\left(\sigma_2+\frac{mu^2}{3k_{\rm B}T_0}\sigma_1\right),
\label{approxD}
\end{equation}
where $\sigma_1$ and $\sigma_2$ are defined in (\ref{sigma12}).
As usual, we leave the dependence on $({\bf r},t)$ implicit.
One easily verifies that (\ref{approxD}) satisfies the constraint (\ref{constD}).
The constant $D$ can be determined by integrating (\ref{eqforD}) over ${\bf p}$:
\begin{eqnarray}
D\int \frac{d{\bf p}}{(2\pi\hbar)^3}\hat L_{12}[\sigma_2+\frac{mu^2}{3k_{\rm B}T_0}\sigma_1]
&=&\frac{1}{\tilde n^{(0)}}\int \frac{d{\bf p}}{(2\pi\hbar)^3}\left(\sigma_2
+\frac{mu^2}{3k_{\rm B}T_0}\right)f_0(1+f_0) \cr
&&-\beta_0 g\tau \int \frac{d{\bf p}}{(2\pi\hbar)^3}\hat L_{12}[1].
\label{eqVC2}
\end{eqnarray}
Using $\hat L_{12}[\frac{mu^2}{2}]=\beta_0 gn_{c0}\hat L_{12}[1]$ and
$\int \frac{d{\bf p}}{(2\pi\hbar)^3}\hat L_{12}[1]=-n_{c0}/\tau_{12}$,
where $\tau_{12}$ is defined in (\ref{tau12}), (\ref{eqVC2}) gives finally
\begin{equation}
D=-\frac{(\tau_{12}+\beta_0 gn_{c0}\tau)}{n_{c0}[\sigma_2+\frac{2}{3}\sigma_1
\beta(\mu_{c0}-U_0)]}.
\label{solutionD}
\end{equation}
Using (\ref{solutionD}) and (\ref{approxD}) in the expression for $\tau$ given by
(\ref{tau}), we can solve to give an explicit expression for $\tau$, namely
\begin{equation}
\frac{1}{\tau}=\frac{1}{\tau_{12}}\left\{\frac{n_{c0}} {\tilde n_0}
\left[\sigma_2+\frac{2}{3}\sigma_1\beta_0(\mu_{c0}-U_0)\right]-\beta_0 g n_{c0}
\right\}
=\frac{\beta_0 gn_{c0}}{\sigma_H \tau_{12}}=\frac{1}{\tau_{\mu}}.
\end{equation}
We thus see that $\tau$ is precisely the relaxation time $\tau_{\mu}$ first
introduced in the ZGN$'$ two-fluid hydrodynamics.
We can now express the four second viscosity coefficients in
(\ref{zeta12}) and (\ref{zeta34}) in terms of the
$\tau_{12}$ collision time defined in (\ref{tau12}):
\begin{equation}
\zeta_1=\frac{k_{\rm B}T}{3m}\sigma_H^2\tau_{12}, ~
\zeta_2=\frac{n_ck_{\rm B}T}{9}\sigma_H^2\tau_{12}, ~
\zeta_3=\frac{k_{\rm B}T}{m^2n_c}\sigma_H^2\tau_{12}, ~
\zeta_4=\zeta_1.
\label{zeta_final}
\end{equation}

\subsection{Numerical results for a uniform Bose gas}
For illustration, we calculate the transport coefficients in (\ref{kappa_final})
and (\ref{etafinal}) for a {\it uniform} Bose gas  ($U_{\rm ext}=0$).
As in Ref.~\cite{CJP}, we choose $gn/k_{\rm B}T_{\rm BEC}=0.2$.
In Fig.1 and Fig.2, we plot the temperature dependence of the dimensionless
transport coefficients $\bar\kappa$ and $\bar\eta$, defined by
\begin{equation}
\bar\kappa\equiv \kappa/nv_{\rm cl}^2\tau_0k_{\rm B},~~
\bar\eta\equiv \eta/nv_{\rm cl}m\tau_0.
\end{equation}
Here $v_{\rm cl}\equiv (5k_{\rm B}T_{\rm BEC}/3m)^{1/2}$ is the hydrodynamic
sound velocity of a classical gas at $T=T_{\rm BEC}$
and $\tau_0^{-1}\equiv \sqrt{2}(8\pi a^2) n (8k_{\rm B}T_{\rm BEC}/\pi m)^{1/2}$
is the classical mean collision time evaluated at $T=T_{\rm BEC}$.
To see the separate effects of the $C_{22}$ collisions and $C_{12}$ collisions,
we also plot the results obtained by taking either $I_{12}^{\kappa,\eta}=0$
or $I_{22}^{\kappa,\eta}=0$.
In Ref.~\cite{CJP}, we neglected $C_{12}$.
We see that the both $\kappa$ and $\eta$ are reduced when we include the $C_{12}$
collisions.
At lower temperatures $T\lesssim 0.5T_{\rm BEC}$, both $\kappa$ and $\eta$ are
dominated by the $C_{12}$ collision integral.

In Fig.~3, we plot the four second viscosity coefficients given in (\ref{zeta_final})
for a uniform Bose gas.
We recall that in Fig.1 of Ref.~\cite{NZG}, we gave the temperature dependence
of $\tau_{\mu}$ and $\tau_{12}$ for $gn/k_{\rm B}T_{\rm BEC}=0.1$.
In Fig.~3, we use the dimensionless second viscosity coefficients, defined by
\begin{equation}
\bar \zeta_1 \equiv \zeta_1/v_{\rm cl}^2\tau_0, ~
\bar \zeta_2 \equiv \zeta_2m/nv_{\rm cl}^2\tau_0, ~
\bar \zeta_3 \equiv \zeta_2n/mv_{\rm cl}^2\tau_0.
\end{equation}

The transport coefficients in a {\it trapped} Bose gas behave quite differently from 
those of a uniform Bose gas.
In particular, since $n_c\gg \tilde n$ holds in the central region of the trap at all
temperatures, the contribution of $C_{12}$ collisions dominates over the 
contribution of the $C_{22}$ collisions at all temperatures below $T_{\rm BEC}$.
We will discuss the implications of this at the end of Section \ref{sec:conclusions}.

\section{conclusions}
\label{sec:conclusions}
In this paper, we have derived two-fluid hydrodynamic equations starting
from the quantum kinetic equation and the generalized GP equation derived in
\cite{ZNG,NZG,milena2}.
However, to complement and extend our earlier work~\cite{ZNG,NZG},
we started from the complete local equilibrium single-particle distribution $f^{(0)}$
as given by (\ref{f0}) and (\ref{muc0}).
Using the Chapman-Enskog approach,
we then included the effects of a small deviation from this local equilibrium form.
This deviation from local equilibrium within the thermal cloud
brings in the usual kind of hydrodynamic
damping due to the thermal conductivity and shear viscosity of the  thermal cloud. 
A summary of our major results is given in the final paragraph of Section \ref{sec:intro}.

In addition,
we showed that the there is additional dissipation associated with the collisional
exchange of atoms between the condensate and non-condensate components.
When we write our hydrodynamic equations in the Landau-Khalatnikov ~\cite{Khal,HM}
form given by (\ref{landau_eqs}), this damping is described in terms of the usual
four second viscosity coefficients for a Bose superfluid.
The appearance of the second viscosity coefficients in the equations for
the total current ${\bf j}$ in (\ref{eq_j}) and for the superfluid velocity ${\bf v}_c$ 
in (\ref{eq_vs}) is due to the deviation of the total pressure and the chemical potential from 
their local equilibrium values.
We might also recall that
Khalatnikov \cite{Khal} discusses a specific model for 
the second viscosity coefficients in superfluid $^4$He by introducing 
``local chemical potentials" for the phonons ($\mu_{\rm ph}$) and rotons ($\mu_{\rm r}$).
These describe a situation where such excitations (describing the normal fluid)
are out of local equilibrium with the superfluid component.
Clearly this discussion has connections with our calculations based on the condensate
and non-condensate not being in diffusive equilibrium.

The frequency-dependence of the second viscosity coefficients is a result of the
fact that our two-fluid hydrodynamics deals with the dynamics of the condensate
and non-condensate components as {\it separate} degrees of freedom.
This feature is made more explicit in our recent papers \cite{ZNG,NZG,CJP}.
In particular, it gives rise to a new relaxational zero frequency mode.
As mentioned at the end of Section~\ref{zgn'}, and more explicitly in Ref.~\cite{CJP},
this mode may be viewed as the (renormalized) version of the zero-frequency thermal
diffusion mode \cite{huang} above $T_{\rm BEC}$.
The presence of this new mode below $T_{\rm BEC}$ is somewhat hidden in the formulation 
in terms of the LK two-fluid equations given in (\ref{landau_eqs}).

In Section \ref{sec:transport},
we derived explicit formulas for all the transport coefficients within our model.
In Ref.~\cite{CJP}, we only took
into account the deviations from local equilibrium due to the 
$C_{22}$ collision integral 
in calculating the shear viscosity and the thermal conductivity coefficients.
In the present paper, we have also included the contribution to these quantities 
from the $C_{12}$ collision integral.
From (\ref{kappa_final2}) and (\ref{etafinal2}), we see that both
$\kappa$ and $\eta$ are given in a form proportional to characteristic relaxation times
$\tau_{\kappa}=(\tau_{\kappa,12}^{-1}+\tau_{\kappa,22}^{-1})^{-1}$ and
$\tau_{\eta}=(\tau_{\eta,12}^{-1}+\tau_{\eta,22}^{-1})^{-1}$, respectively,
which are defined and motivated in Appendix \ref{relax_times}.

In a rough estimate, we find $\tau_{\kappa,22},\tau_{\eta,22} \sim \tau_{\rm cl}
\sim 1/\tilde n$
and $\tau_{\kappa,12},\tau_{\eta,12}\sim (\tilde n/ n_c) \tau_{\rm cl}
\sim 1/n_c$,
where $\tau_{\rm cl}$ is the classical collision time defined in
(\ref{taucl}).
We therefore observe that the effect of $C_{12}$ collisions reduces the magnitude
of both $\kappa$ and $\eta$ by a factor $\sim 1/(1+n_c/\tilde n)$,  a result also
noted in Ref.~\cite{KD} for a {\it uniform} gas.
The contribution of the $C_{12}$ collisions is always important in a trapped gas,
since the condensate density at the central region of 
a trap is much larger than the density of the thermal cloud even at temperatures
very close to $T_{\rm BEC}$.
In a trapped Bose gas, we find the effect of the $C_{12}$ collisions is enhanced by
a large factor $n_c/\tilde n \gg 1$.
This means that $\kappa$ and $\eta$ are always dominated by the contribution of the
$C_{12}$ collisions.
Since the  effective relaxation times $\tau_{\kappa,12}$ and
$\tau_{\eta,12}$ are smaller than the classical collision time $\tau_{\rm cl}$
by a factor $\tilde n/n_c\ll 1$, this
implies that in a trapped Bose gas, the hydrodynamic region may be {\it much easier}
to reach at finite temperatures than expected
from naive considerations based on using the classical collision time
(i.e., $\omega\tau_{\rm cl}\ll 1$).
That is to say, one might easily have $\omega\tau_{\kappa,12}\ll 1$ and
$\omega\tau_{\eta,12}\ll 1$, even though $\omega\tau_{\rm cl}\gg 1$.
This has very important implications in deciding if one is in the
collisionless or the hydrodynamic region.

One problem not dealt with in this paper is the fact that in a trapped Bose gas,
the decreasing density in the tail of the thermal cloud means that the hydrodynamic
description breakdowns eventually.
This problem enters the evaluation of the expressions for the
$\eta$ and $\kappa$ transport coefficients given in Section \ref{sec:transport}.
In recent papers dealing with the case above $T_{\rm BEC}$ \cite{KPS,NG},
this problem was handled in an ad-hoc manner by introducing a spatial cutoff in the 
thermal cloud.
In a future paper, we given an alternative approach based on starting with an
improved solution of the kinetic equation, which naturally includes the cross-over to the
non-interacting gas in the thermal gas tail.

\begin{center}
{\bf ACKNOWLEDGMENTS}
\end{center}
We thank E. Zaremba for his interest and comments and G. Shlyapnikov for a
useful discussion.
T.N. was supported by JSPS of Japan, while A.G. was supported by NSERC
of Canada.

\begin{appendix}
\section{}

We briefly sketch the derivation of the first-order kinetic equation given in (\ref{linpsi}).		Using (\ref{f0}) in (\ref{eq_f1}), one has
\begin{eqnarray}
&&\left[ \frac{\partial^0}{\partial t} +\frac{{\bf p}}{m}\cdot
\bbox{\nabla}_{{\bf r}} -\bbox{\nabla}_{{\bf r}} U({\bf r},t)\cdot\bbox{\nabla}_{\bf p}\right]
f^{(0)}({\bf r},{\bf p},t) \nonumber \\
&&=\left[\frac{1}{z}\left(\frac{\partial^0}{\partial t}+
\frac{{\bf p}}{m}\cdot\bbox{\nabla} \right)z
+\frac{mu^2}{2k_{\rm B}T^2}
\left(\frac{\partial^0}{\partial t}+
\frac{{\bf p}}{m}\cdot\bbox{\nabla} \right)T \right. \nonumber \\
&&\ \ \ \  \left. +\frac{m{\bf u}}{k_{\rm B}T}\cdot
\left(\frac{\partial^0}{\partial t}+
\frac{{\bf p}}{m}\cdot\bbox{\nabla} \right){\bf v}_n
+\frac{\bbox{\nabla}U({\bf r},t)}{k_{\rm B}T}\cdot {\bf u}
\right] f^{(0)}(1+f^{(0)}).
\label{A1}
\end{eqnarray}
The notation $\partial^0/\partial t$ is explained below (\ref{eq_f1}).
Using the expressions for the density $\tilde n^{(0)}$
given by (\ref{ntilde0})
and the pressure $\tilde P^{(0)}$ in (\ref{ptilde0}), one finds
\begin{eqnarray}
\frac{\partial^0\tilde n^{(0)}}{\partial t}
&=&\frac{3\tilde n^{(0)}}{2 T}\frac{\partial^0 T}{\partial t}
+\frac{\gamma k_{\rm B}T}{z}\frac{\partial^0 z}{\partial t} \,, \cr
&& \cr
\frac{\partial^0 \tilde P^{(0)}}{\partial t}
&=&\frac{5\tilde P^{(0)}}{2T}\frac{\partial^0 T}{\partial t}
+\frac{\tilde n^{(0)}k_{\rm B}T}{z}\frac{\partial^0 z}{\partial t}\,,
\end{eqnarray}
where $\gamma$ is the variable introduced after (\ref{sigma12})
and $z=z^{(0)}$ as defined below (\ref{ntilde0}).
One may combine these equations with (\ref{eq_ntilde0}) and (\ref{eq_ptilde0})
to show that the equations in (A2) reduce to
\begin{eqnarray}
\frac{\partial^0 T}{\partial t}&=&-\frac{2}{3}T(\bbox{\nabla}\cdot{\bf v}_n)
-{\bf v}_n\cdot\bbox{\nabla}T+\frac{2T}{3\tilde n^{(0)}}\sigma_1 \Gamma_{12}, \cr
&& \cr
\frac{\partial^0 z}{\partial t}&=&-{\bf v}_n\cdot\bbox{\nabla}z+
\sigma_2z\frac{\Gamma_{12}}{\tilde n^{(0)}}.
\end{eqnarray}
The analogous equation for $\partial^0{\bf v}_n/\partial t$ is given directly
by (\ref{eq_vn0}).
Using these results in (A1), one finds that it reduces to
\begin{eqnarray}
&&\left(\frac{\partial^0}{\partial t}
+\frac{{\bf p}}{m}\cdot\bbox{\nabla}_{{\bf r}}-\bbox{\nabla}_{{\bf r}}U\cdot\bbox{\nabla}_{\bf p}
\right)f^{(0)} \cr
&&=\left\{
\frac{1}{T}{\bf u}\cdot\bbox{\nabla}T\left(\frac{mu^2}{2k_{\rm B}T}
-\frac{5\tilde P^{(0)}}{2\tilde n^{(0)}k_{\rm B}T} \right)
+\frac{m}{k_{\rm B}T}\left[{\bf u}\cdot({\bf u}\cdot\bbox{\nabla}){\bf v}_n
-\frac{u^2}{3}\bbox{\nabla}\cdot{\bf v}_n\right]\right. \cr
&&+\left.\left[\sigma_2+\frac{mu^2}{3k_{\rm B}T}\sigma_1+\frac{m}{k_{\rm B}T}
{\bf u}\cdot({\bf v}_c-{\bf v}_n) \right]
\frac{\Gamma_{12}}{\tilde n^{(0)}}\right\}f^{(0)}
(1+f^{(0)}),
\label{lhs_f1}
\end{eqnarray}
where we recall ${\bf u}\equiv {\bf p}/m-{\bf v}_n$.
In calculating the dissipative terms, we only consider terms to
first order in the velocity fields ${\bf v}_n$ and ${\bf v}_c$.
Since $\Gamma_{12}$ is proportional to ${\bf v}_n$ and ${\bf v}_c$
(see (\ref{gamma1})),
we can neglect the last term (proportional to ${\bf v}_c-{\bf v}_n$) in (\ref{lhs_f1}).
This linearized version of (\ref{lhs_f1}) 
can be rewritten in the form shown on the left hand side of (\ref{linpsi}).

\section{}
\label{relax_times}
The relaxation time $\tau_{\kappa}$ in (\ref{kappa_final2}) is defined by
\begin{eqnarray}
\tau_{\kappa}&\equiv& -
\left\{\int \frac{d{\bf p}}{(2\pi\hbar)^3}
\left[\frac{mu^2}{2k_{\rm B}T_0}
-\frac{5g_{5/2}(z_0)}{2g_{3/2}(z_0)}\right]^2u^2
f_0(1+f_0) \right\} \cr
&& \times \left\{\int \frac{d{\bf p}}{(2\pi\hbar)^3} 
\left[\frac{mu^2}{2k_{\rm B}T_0}
-\frac{5g_{5/2}(z_0)}{2g_{3/2}(z_0)}\right]{\bf u} \cdot
\hat L \left[\left\{\frac{mu^2}{2k_{\rm B}T_0}
-\frac{5g_{5/2}(z_0)}{2g_{3/2}(z_0)} \right\}{\bf u}\right]\right\}^{-1} \cr
~~&& \cr 
&=& \frac{ 15\pi^{9/2}\hbar^3 }{ 8m(k_{\rm B}T_0)^2a^2 }
\left[\frac{ \frac{7}{2}g_{7/2}(z_0)-\frac{5}{2}g_{5/2}^2(z_0)/g_{3/2}(z_0) }
{I_{22}^{\kappa}(z_0)+n_{c0}\Lambda_0^3 I_{12}^{\kappa}(z_0) } \right]\cr
~~&& \cr
&=&
\frac{15\sqrt{2}\pi^{7/2}}{4}\tau_{\rm cl}
\left[\frac{ \frac{7}{2}g_{7/2}(z_0)g_{3/2}(z_0)-\frac{5}{2}g_{5/2}^2(z_0) }
{I_{22}^{\kappa}(z_0)+n_{c0}\Lambda_0^3 I_{12}^{\kappa}(z_0) }\right].
\label{tau_k}
\end{eqnarray}
Here 
\begin{equation}
\tau_{\rm cl}^{-1}\equiv \sqrt{2}(8\pi a^2) \tilde n_0({\bf r})
 (8k_{\rm B}T_0/\pi m)^{1/2},
\label{taucl}
\end{equation}
is the collision time of a classical gas with density $\tilde n_0$ and quantum 
cross section $\sigma=8\pi a^2$.
In turn, the relaxation times in (\ref{tau_k2}) are defined by
\begin{eqnarray}
\tau_{\kappa,22}&\equiv& - \left\{\int \frac{d{\bf p}}{(2\pi\hbar)^3}
\left[\frac{mu^2}{2k_{\rm B}T_0}-\frac{5g_{5/2}(z_0)}{2g_{3/2}(z_0)}\right]^2u^2
f_0(1+f_0) \right\}\cr
&& \times \left\{\int \frac{d{\bf p}}{(2\pi\hbar)^3} 
\left[\frac{mu^2}{2k_{\rm B}T_0}
-\frac{5g_{5/2}(z_0)}{2g_{3/2}(z_0)}\right]{\bf u} \cdot
\hat L_{22} \left[\left\{\frac{mu^2}{2k_{\rm B}T_0}
-\frac{5g_{5/2}(z_0)}{2g_{3/2}(z_0)} \right\}{\bf u}\right] \right\}^{-1} \cr
~~&& \cr 
&=& \frac{ 15\pi^{9/2}\hbar^3 }{ 8m(k_{\rm B}T_0)^2a^2 }
\left[\frac{ \frac{7}{2}g_{7/2}(z_0)-
\frac{5}{2}g_{5/2}^2(z_0)/
g_{3/2}(z_0) }{I_{22}^{\kappa}(z_0)} \right]\cr
~~&& \cr
&=&\frac{15\sqrt{2}\pi^{7/2}}{4}\tau_{\rm cl}
\left[\frac{ \frac{7}{2}g_{7/2}(z_0)g_{3/2}(z_0)-
\frac{5}{2}g_{5/2}^2(z_0)}
{I_{22}^{\kappa}(z_0)}\right],
\end{eqnarray}
and
\begin{eqnarray}
\tau_{\kappa,12}&\equiv&
 - \left\{\int \frac{d{\bf p}}{(2\pi\hbar)^3}
\left[\frac{mu^2}{2k_{\rm B}T_0}-\frac{5g_{5/2}(z_0)}{2g_{3/2}(z_0)}\right]^2u^2
f_0(1+f_0) \right\}
\cr &&
 \times \left\{ \int \frac{d{\bf p}}{(2\pi\hbar)^3} 
\left[\frac{mu^2}{2k_{\rm B}T_0}
-\frac{5g_{5/2}(z_0)}{2g_{3/2}(z_0)}\right]{\bf u} \cdot
\hat L_{12} \left[\left\{\frac{mu^2}{2k_{\rm B}T_0}
-\frac{5g_{5/2}(z_0)}{2g_{3/2}(z_0)} \right\}{\bf u}\right] \right\}^{-1} \cr
~~&&
 \cr 
&=& \frac{ 15\pi^{9/2}\hbar^3 }{ 8m(k_{\rm B}T_0)^2a^2 }
\left[\frac{ \frac{7}{2}g_{7/2}(z_0)-
\frac{5}{2}g_{5/2}^2(z_0)/
g_{3/2}(z_0) }
{n_{c0}\Lambda_0^3 I_{12}^{\kappa}(z_0) } \right]\cr
~~&& \cr
&=&
\frac{15\sqrt{2}\pi^{7/2}} {4} \tau_{\rm cl}
\left[ \frac{ \frac{7}{2}g_{7/2}(z_0)g_{3/2}(z_0)-\frac{5}{2}g_{5/2}^2(z_0)}
{n_{c0}\Lambda_0^3 I_{12}^{\kappa}(z_0) } \right].
\end{eqnarray}

In an analogous way, the
relaxation time $\tau_{\eta}$ in (\ref{etafinal2}) is defined by
\begin{eqnarray}
\tau_{\eta}&\equiv&
-\left[\int \frac{d{\bf p}}{(2\pi\hbar)^3}
\left(u_{\mu}u_{\nu}-\frac{1}{3}\delta_{\mu\nu}u^2\right)^2
f_0(1+f_0) \right]\cr
&&\times\left\{
\int \frac{d{\bf p}}{(2\pi\hbar)^3}\left(
u_{\mu}u_{\nu}-\frac{1}{3}\delta_{\mu\nu}u^2
\right) \hat L\left[u_{\mu}u_{\nu}-\frac{1}{3}\delta_{\mu\nu}u^2\right]
\right\}^{-1} \cr
~~&& \cr
&=&\frac{5\pi^{9/2}\hbar^3}{16m(k_{\rm B}T_0)^2a^2}
\left[\frac{ g_{5/2}(z_0) }{I_{22}^{\eta}(z_0)+n_{c0}\Lambda_0^3 I_{12}^{\eta}(z_0)}\right]
\cr ~~&& \cr
&=&\frac{5\sqrt{2}\pi^{7/2}}{2}\tau_{\rm cl}
\left[\frac{g_{5/2}(z_0)g_{3/2}(z_0)}
{I_{22}^{\eta}(z_0)+n_{c0}\Lambda_0^3I_{12}^{\eta}(z_0)} \right].
\label{tau_e}
\end{eqnarray}
The relaxation times in (\ref{tau_e2}) are defined by
\begin{eqnarray}
\tau_{\eta,22}&\equiv&
-\left[\int \frac{d{\bf p}}{(2\pi\hbar)^3}
\left(u_{\mu}u_{\nu}-\frac{1}{3}\delta_{\mu\nu}u^2\right)^2
f_0(1+f_0) \right]\cr
&&\times\left\{
\int \frac{d{\bf p}}{(2\pi\hbar)^3}\left(
u_{\mu}u_{\nu}-\frac{1}{3}\delta_{\mu\nu}u^3
\right) \hat L_{22}\left[u_{\mu}u_{\nu}-\frac{1}{3}\delta_{\mu\nu}u^3\right]
\right\}^{-1} \cr ~~ && \cr
&=&\frac{5\pi^{9/2}\hbar^3}{16m(k_{\rm B}T_0)^2a^2}
\left[\frac{ g_{5/2}(z_0) }{I_{22}^{\eta}(z_0)}\right] \cr ~~&& \cr
&=&\frac{5\sqrt{2}\pi^{7/2}}{2}\tau_{\rm cl}
\left[\frac{g_{5/2}(z_0)g_{3/2}(z_0)}
{I_{22}^{\eta}(z_0)} \right],
\end{eqnarray}
and
\begin{eqnarray}
\tau_{\eta,12}&\equiv&
-\left[\int \frac{d{\bf p}}{(2\pi\hbar)^3}
\left(u_{\mu}u_{\nu}-\frac{1}{3}\delta_{\mu\nu}u^2\right)^2
f_0(1+f_0) \right]\cr
&&\times\left\{
\int \frac{d{\bf p}}{(2\pi\hbar)^3}\left(
u_{\mu}u_{\nu}-\frac{1}{3}\delta_{\mu\nu}u^2
\right) \hat L_{12}\left[u_{\mu}u_{\nu}-\frac{1}{3}\delta_{\mu\nu}u^2\right]
\right\}^{-1} \cr ~~ && \cr
&=&\frac{5\pi^{9/2}\hbar^3}{16m(k_{\rm B}T_0)^2a^2}
\left[\frac{ g_{5/2}(z_0) }{n_{c0}\Lambda_0^3 I_{12}^{\eta}(z_0)}\right] \cr 
~~&& \cr
&=&\frac{5\sqrt{2}\pi^{7/2}}{2}\tau_{\rm cl}\left[
\frac{g_{5/2}(z_0)g_{3/2}(z_0)}{n_{c0}\Lambda_0^3I_{12}^{\eta}(z_0)}
\right].
\end{eqnarray}

We note that in a non-degenerate gas, these expressions simplify and we find
$\tau_{\kappa}=\tau_{\kappa,22}=\frac{15}{8}\tau_{\rm cl}(n_0({\bf r}))$
and
$\tau_{\eta}=\tau_{\eta,22}=\frac{5}{4}\tau_{\rm cl}(n_0({\bf r}))$.
The latter expression for $\tau_{\eta}$ agree with the result for the
shear viscosity given in Eq.~(11) by Kavoulakis et al. \cite{KPS2}.

The physical meaning of these new relaxation times becomes clear when we compare
our Chapman-Enskog analysis in the text with a simple relaxation time approximation \cite{huang}.
In the relaxation time approximation, one simply replaces the collision term in (\ref{eq1}) with
$-[f-f^{(0)}]/\tau_{\rm rel}$, where $\tau_{\rm rel}$ is a phenomenological
relaxation time characterizing how fast the system relaxes to local equilibrium.
With this approximation, the solutions of the linearized equations in
(\ref{eqforA}) and (\ref{eqforB}) for the functions $A(u)$ and $B(u)$
are found to be simply given by
\begin{equation}
A(u)=-\tau_{\rm rel}\left[\frac{mu^2}{2k_{\rm B}T}-\frac{5g_{5/2}(z)}
{2g_{3/2}(z)}\right], ~~
B(u)=-\tau_{\rm rel}\frac{m}{k_{\rm B}T}.
\label{relax_approx}
\end{equation}
In contrast, our Chapman-Enskog solution for $A(u)$ is given by (\ref{ansatzA})
with the coefficient $A$ given by (\ref{eqAconst}), while one has
$B(u)=B$ with the constant $B$ given by (\ref{eqBconst}).
In terms the relaxation times $\tau_{\kappa}$ and $\tau_{\eta}$ defined above, 
we find (\ref{ansatzA}) and (\ref{eqBconst}) can be written as
\begin{equation}
A(u)=-\tau_{\kappa}\left[\frac{mu^2}{2k_{\rm B}T}-\frac{5g_{5/2}(z)}
{2g_{3/2}(z)}\right], ~~
B(u)=-\tau_{\eta}\frac{m}{k_{\rm B}T}.
\label{relax_approx2}
\end{equation}
Comparing (\ref{relax_approx2}) with (\ref{relax_approx}), we see that
both $\tau_{\kappa}$ and $\tau_{\eta}$ can be identified with the relaxation time
$\tau_{\rm rel}$.
That is, in the simple relaxation time approximation, $\kappa$ and $\eta$
are still given by the formulas (\ref{kappa_final2}) and (\ref{etafinal2}),
but with $\tau_{\kappa}=\tau_{\eta}=\tau_{\rm rel}$.
We also note that Eq.~(7) of Ref.~\cite{KPS2} gives a general expression for various 
collisional relaxation times above $T_{\rm BEC}$, which is given by 
a formula analogous to (\ref{tau_k}) and (\ref{tau_e}).

It is these effective relaxation times that determine whether one is in the hydrodynamic
two-fluid domain, $\omega\tau_{\kappa}\lesssim 1$, $\omega\tau_{\eta}\lesssim 1$
(see discussion in Section \ref{sec:conclusions}).
They should be used in place of the classical collision time $\tau_{\rm cl}$ in
(\ref{taucl}). 

\section{}

Hohenberg and Martin~\cite{HM} worked out the dispersion relation of the
hydrodynamic modes in a
a {\it uniform} Bose superfluid using the Landau-Khalatnikov two-fluid equations.
The frequencies of the first and second sound modes are 
given by
\begin{equation}
\omega^2=u_i^2k^2-iD_ik^2\omega,
\end{equation}
where the sound velocities $u_i$ are determined by the coupled equations
\begin{eqnarray}
u_1^2+u_2^2&=&\frac{T\rho_s {\bar s}^2}{\rho_n {\bar c_v}}+\left.\frac{\partial P}
{\partial {\bar s}}\right|_{\bar{s}}, \\
u_1^2u_2^2&=&\frac{T\rho_s{\bar s}^2}{\rho_n\bar{c_v}}\left.
\frac{\partial P}{\partial \rho}\right|_{T}.
\end{eqnarray}
Here $\bar s\equiv s/mn$ is the entropy per unit mass and $\bar c_v$ is the
specific heat per unit mass.
The damping coefficients $D_i$ are determined by the coupled equations
\begin{eqnarray}
D_1+D_2&=&\frac{4\eta}{3\rho_n}+\frac{\zeta_2}{\rho_n}-\frac{\rho_s}{\rho_n}
(\zeta_1+\zeta_4)+\frac{\zeta_3\rho_s}{\rho_n}\rho+\frac{\kappa}{\rho {\bar c_v}}, \\
u_1^2D_1+u_2^2D_2&=&\frac{\partial P}{\partial \rho}+\frac{\zeta_2+4\eta /3 }{\rho}
\frac{\rho_s}{\rho_n}\left(\frac{T{\bar s}^2}{\bar c_v}-\frac{2T{\bar s}}
{\rho {\bar c_v}} \left.\frac{\partial P}{\partial T}\right|_{\rho}
+\left.\frac{\partial P}{\partial \rho} \right|_{\bar s}\right)
+\zeta_3\frac{\rho_s}{\rho_n}\left. \frac{\partial P}{\partial \rho}\right|_{\bar s}\rho 
\\
&&-(\zeta_1+\zeta_4)\frac{\rho_s}{\rho_n}\left(-\frac{T{\bar s}}{{\bar c_v} \rho}
\left.\frac{\partial P}{\partial T}\right|_{\rho}+\left.\frac{\partial P}{\partial\rho}
\right|_{\bar s}\right).
\end{eqnarray}

We note that the above general expressions are valid for both liquid $^4$He and
Bose gases.
For the liquid $^4$He, these formulas can be simplified by using ${\bar c_v}\approx 
{\bar c_p}$. 
One has
\begin{eqnarray}
u_1^2&\approx&\frac{\partial P}{\partial \rho}  \\
u_2^2&\approx&-\frac{\rho_s}{\rho_n}\frac{\partial T}{\partial (1/{\bar s})}, \\
D_1&\approx&\frac{\zeta_2+4\eta/3}{\rho}, \\
D_2&\approx&\frac{1}{\rho}
\left\{ \frac{\kappa}{ T(\partial{\bar s}/\partial T)}+\frac{\rho_s}{\rho_n}
\left[\zeta_3\rho^2-\rho(\zeta_1+\zeta_4)+\zeta_2+\frac{4}{3}\eta\right]\right\}.
\end{eqnarray}
However, these are not valid for a dilute Bose gas.
Calculating the various thermodynamic derivatives with the
Hartree-Fock single-particle spectrum used in this paper~\cite{GZ},
the sound velocities and damping coefficients for a uniform Bose gas are given by
\begin{eqnarray}
u_1^2&\approx& \frac{5\tilde P_0}{3m\tilde n_0}+\frac{2g\tilde n_0}{m}
\left(1-\frac{2n_{c0}^2}{9\tilde n_0^2} \sigma_H \right) \\
u_2^2&\approx& \frac{gn_{c0}}{m}(1-\sigma_H)\\
D_1&\approx& \frac{4\eta}{3\rho_n}+\frac{\zeta_2}{\rho_n}
-\frac{\rho_s}{\rho_n}
(\zeta_1+\zeta_4)+\frac{\zeta_3\rho_s^2}{\rho_n}
+\frac{4}{9}\frac{\kappa T_0}{\rho_n u_1^2}\left(1+\frac{2n_{c0}}{3\tilde n_0}
\sigma_1\sigma_H \right)^2, 
\\
D_2&\approx&\rho_s\zeta_3+\frac{4}{9}\frac{\rho_s}{\rho_n^2 u_2^2}(\sigma_1\sigma_H)^2
\kappa T_0 .
\end{eqnarray} 
Here $\sigma_H$ is defined in (\ref{sigma_H}), while $\rho_n=m\tilde n_0$ and $\rho_s=mn_{c0}$.

\end{appendix}


\centerline{\bf FIGURE CAPTIONS}
\begin{itemize}
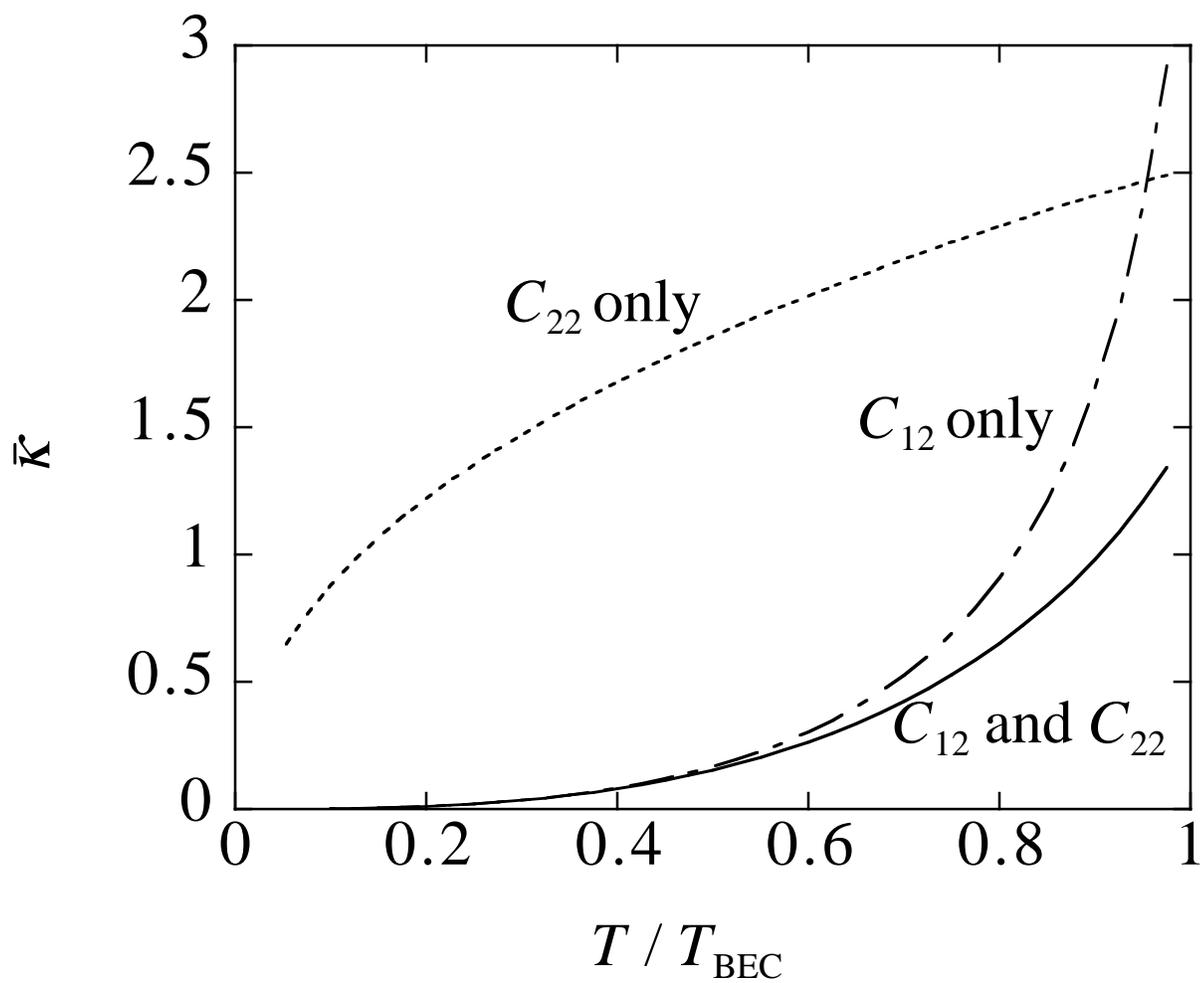
\item[FIG.1:] 
Plot of the thermal conductivity $\kappa$ in a uniform gas for 
$gn=0.2k_{\rm B}T_{\rm BEC}$ as a function of temperature.
We also plot the results by taking $I_{12}^{\kappa}=0$ (dashed line)
and $I_{22}^{\kappa}=0$ (dot-dashed line).

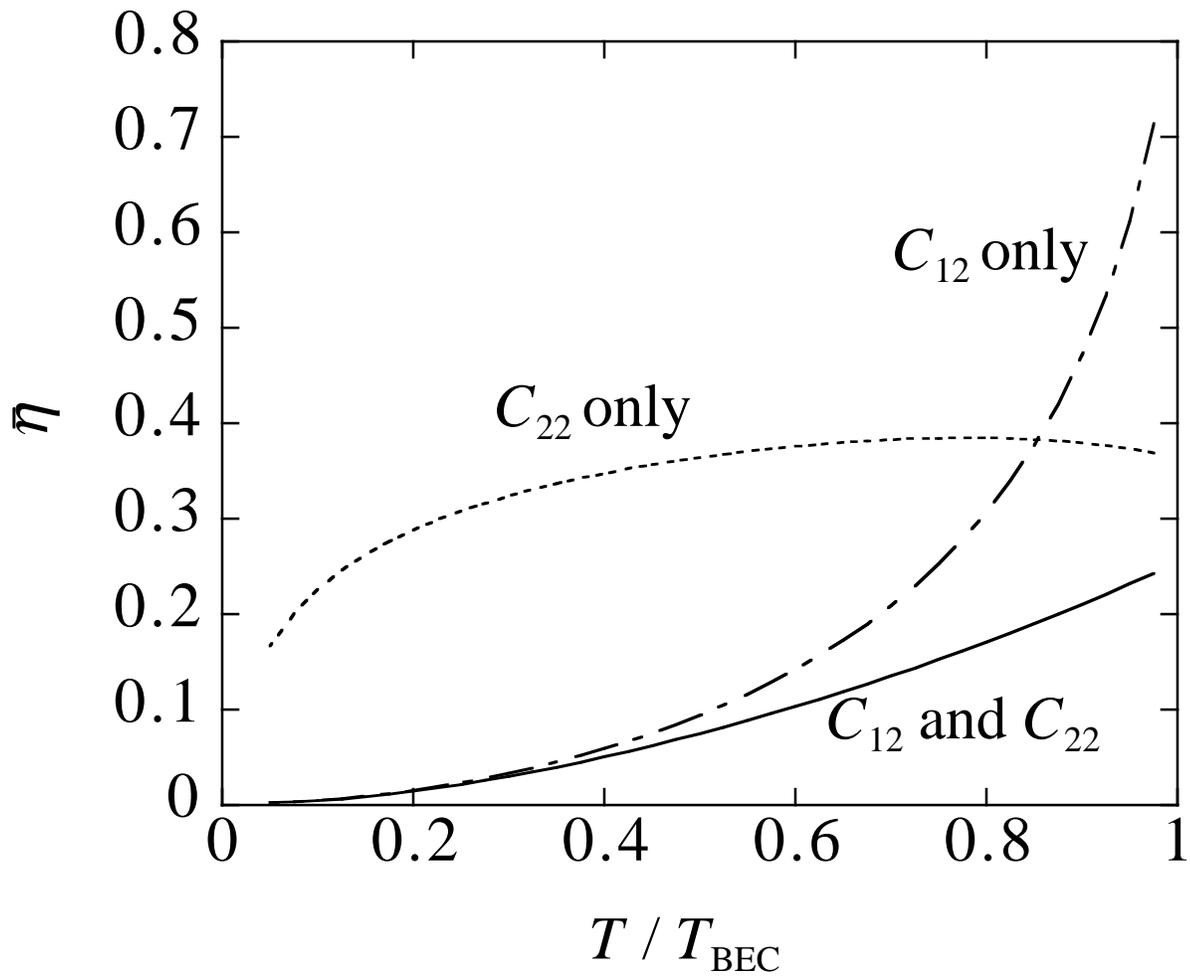
\item[FIG.2:] 
Plot of the shear viscosity coefficient in a uniform gas for 
$gn=0.2k_{\rm B}T_{\rm BEC}$ as a function of temperature.
We also plot the results by taking $I_{12}^{\eta}=0$ (dashed line)
and $I_{22}^{\eta}=0$ (dot-dashed line).

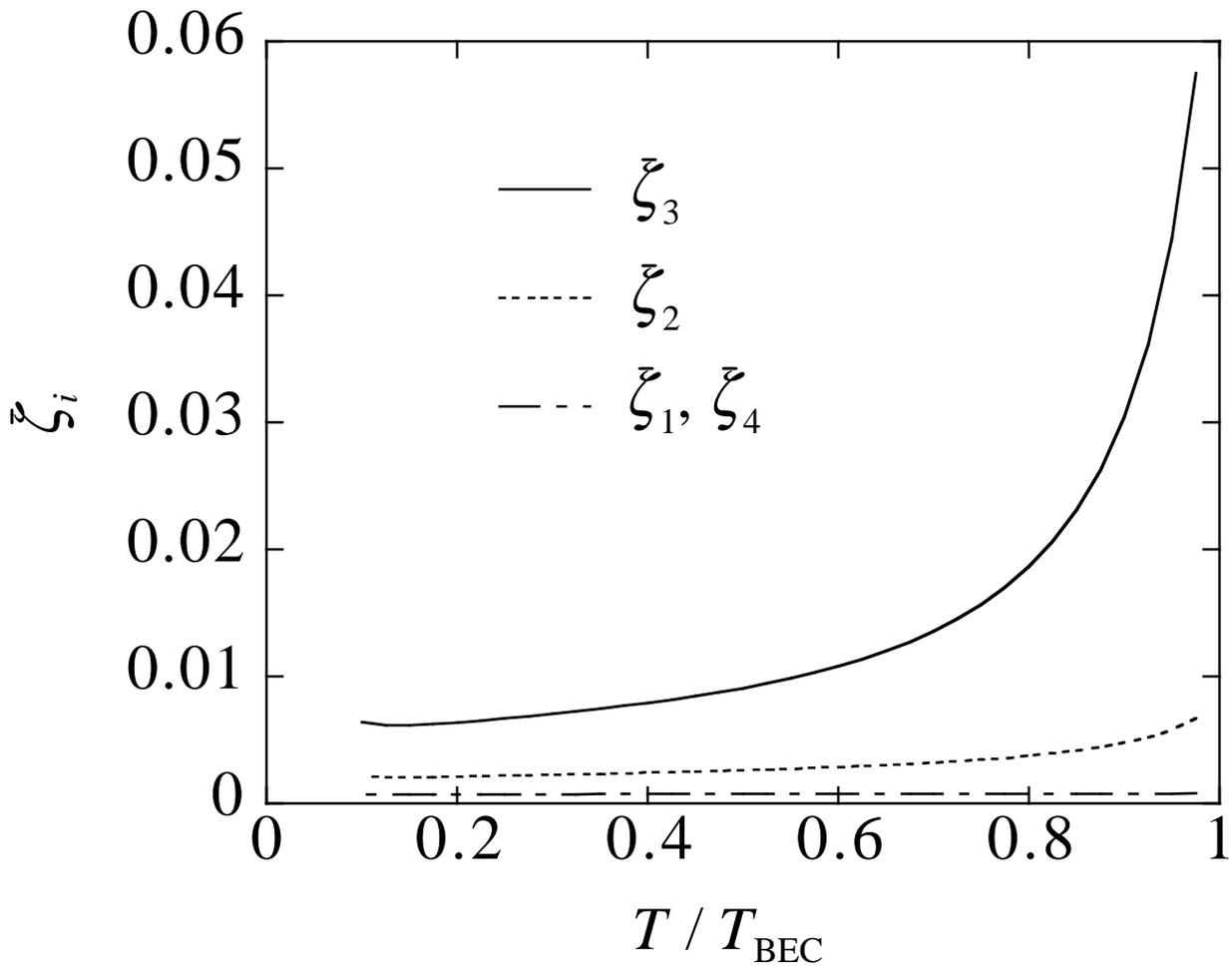
\item[FIG.3:] 
Plot of the second viscosity coefficients $\zeta_i$
in a uniform gas for $gn=0.2k_{\rm B}T_{\rm BEC}$
as a function of temperature.

\end{itemize}
\end{document}